\documentclass[final,3p,times,12pt,fixfloat]{elsarticle}
\journal{Nuclears Physics B}

\bibpunct{[}{]}{,}{n}{}{,} 

\usepackage{graphicx}
\usepackage{subfig}

\usepackage{multirow}
\usepackage{amsmath}
\usepackage{bm}
\usepackage{comment}
\usepackage{txfonts}
\usepackage{tabularx,ragged2e}%

\newcolumntype{Y}{>{\centering\arraybackslash}X}

\usepackage{lineno}

\newcommand{\beq}{\begin{eqnarray}}
\newcommand{\eeq}{\end{eqnarray}}
\newcommand{\be}{\begin{eqnarray*}}
\newcommand{\ee}{\end{eqnarray*}}

\newcommand{\vect}[1]{\overset{\to}{#1}}
\renewcommand{\vec}[1]{\vect{#1}}

\newcommand{\kks}[1]{#1 \!\!\! \slash }

\newcommand{\eg}{{\it e.g.}}

\newcommand{\lsim}{\mathrel{\vcenter
{\hbox{$<$}\nointerlineskip\hbox{$\sim$}}}}
\newcommand{\gsim}{\mathrel{\vcenter
{\hbox{$>$}\nointerlineskip\hbox{$\sim$}}}}

\def\lsim{\raise0.3ex\hbox{$<$\kern-0.75em\raise-1.1ex\hbox{$\sim$}}}
\def\gsim{\raise0.3ex\hbox{$>$\kern-0.75em\raise-1.1ex\hbox{$\sim$}}}

\def\beq     {\begin{equation}}
\def\eeq     {\end{equation}}

\def\beq     {\begin{equation}}
\def\eeq     {\end{equation}}

\usepackage{ifpdf}
\ifpdf
\usepackage[pdftex]{hyperref}
\else
\usepackage[hypertex]{hyperref}
\fi

\hypersetup{
  pdftitle={Probing unpolarised and linearly polarised gluon TMDs through  $|Upsilon+Z$ 
production at the LHC},%
  pdfauthor={J.P. Lansberg,C. Pisano,M. Schlegel},%
  pdfsubject={},%
  pdfkeywords={Quarkonium Production,  Gluon TMDs, Large Hadron Collider},%
  pdfstartview={},%
  bookmarksopen=true, breaklinks=true, debug=true, %
  colorlinks=true, linkcolor=red, citecolor=blue, urlcolor=blue
}

\begin{document}

\begin{frontmatter}

\title{Associated production of a dilepton and a $\Upsilon(J/\psi)$ at the LHC as a probe of  
gluon transverse momentum dependent distributions}

\author[IPNO]{Jean-Philippe Lansberg}
\author[PAVIA1,PAVIA2]{Cristian Pisano}
\author[TUEBINGEN]{Marc Schlegel}
\address[IPNO]{IPNO, CNRS/IN2P3, Univ. Paris-Sud, Universit\'e Paris-Saclay, 91406 Orsay, France}
\address[PAVIA1]{Dipartimento di Fisica, Universit\`a di Pavia, via Bassi 6, I-27100 Pavia, Italy}
\address[PAVIA2]{INFN Sezione di Pavia, via Bassi 6, I-27100 Pavia, Italy}
\address[TUEBINGEN]{Institute for Theoretical Physics, Universit\"{a}t T\"{u}bingen,
  Auf der Morgenstelle 14, D-72076 T\"{u}bingen, Germany}

\date{\today}

\begin{abstract}
We discuss the impact on the study of gluon transverse momentum dependent distributions (TMDs) of the associated production of a lepton pair and a $\Upsilon$ (or a $J/\psi$) in unpolarised proton-proton collisions, $pp\to {\cal Q} \, \ell \bar{\ell} X$, at LHC energies, where one can assume that such final states are dominantly induced by gluon fusion. If the transverse momentum of the quarkonium-dilepton system -- namely, the transverse momentum imbalance of the quarkonium state and the lepton pair -- is small, the corresponding cross sections can be calculated within the framework of TMD factorisation. Using the helicity formalism, we show in detail how these cross sections are connected to the moments of two independent TMDs: the distribution of unpolarised gluons, $f_1^g$, and the distribution of linearly polarised gluons, $h_1^{\perp g}$. We complete our exhaustive derivation of these general relations with a phenomenological analysis of the feasibility of the TMD extraction,  as well as some outlooks.  
\end{abstract}

\end{frontmatter}

\section{Introduction}
\label{intro}

Three-dimensional momentum distributions of gluons in the nucleon -- the so-called gluon transverse momentum dependent distributions (TMDs) -- have attracted much attention recently~\cite{Mulders:2000sh,Boer:2015vso}. Theoretically, gluon TMDs appear in factorisation formulae that explicitly take the transverse momentum of gluons  into account (TMD factorisation), see \eg~\cite{Collins:2011zzd,Aybat:2011zv,GarciaEchevarria:2011rb,Ma:2012hh}). These formulae apply to cross sections that are differential in the transverse momentum of the final state, ${\bm q}_T$, in a kinematical region where it is much smaller than the hard scale of the process $Q$, $|{\bm q}_T|\ll Q$. Typically, the hard scale refers to the virtuality of the exchanged gauge boson in lepton-nucleon collisions or the invariant mass of the final state  in hadron-hadron collisions.  Our understanding of the evolution with the hard scale of the gluon TMDs has recently significantly improved, see \eg\ \cite{Sun:2011iw,Boer:2014tka,Echevarria:2015uaa,Echevarria:2012pw}.

Several processes have been identified as sensitive probes of gluon TMDs. Probably, the theoretically cleanest one is the production of a pair of (almost) back-to-back heavy quark and antiquark or of a di-jet in lepton-nucleon collisions~\cite{Boer:2010zf,Pisano:2013cya,Boer:2016fqd}. A measurement of such processes could however only be performed at a future Electron-Ion-Collider. On the other hand, reactions initiated by two protons can also provide insights on the gluon TMDs at the LHC or at RHIC.  For instance,  one possibility is to look at di-photon production, still with a large azimuthal separation~\cite{Qiu:2011ai}. This process, however, suffers from additional contributions from quark-induced channels, at RHIC energies in particular. As such, it is probably not the cleanest probe of gluon TMDs in hadron collisions that one could think of. Furthermore, the experimental detection of direct photons requires a specific isolation procedure which may be difficult to implement in a realistic measurement. 

A handier probe of gluon TMDs is certainly to be found among the production of quarkonium states~\cite{Boer:2012bt,Lansberg:2014myg,Angeles-Martinez:2015sea,Lansberg:2015hla,Signori:2016jwo,Signori:2016lvd,Boer:2016bfj}, built up of either charm or bottom quarks, since they are often dominantly produced through gluon fusion at proton-proton colliders and some of them, like the spin triplet vector states, are relatively easy to detect in their di-muons channels.  As for now, the access to gluon TMDs has been investigated in~\cite{Boer:2012bt,Ma:2012hh} with single inclusive $\eta_{c,b}$ or $\chi_{c,b}$-production in proton collisions. One drawback of such single-particle analyses is that they are restricted to low transverse momenta, typically below half the quarkonium mass, which makes such experimental studies particularly challenging. Furthermore, in the case of $\chi_{c,b}$-production~\cite{Ma:2014oha}, TMD factorisation may not hold because of infrared divergences specific to the $P$-wave production\footnote{For review, the readers is referred to ~\cite{Andronic:2015wma,Brambilla:2010cs,Lansberg:2006dh}}. Such issues do not appear for spin singlet $S$-wave state production, which however has so far only been studied for the $\eta_c$ down to $P_T \simeq 6$ GeV by the LHCb collaboration~\cite{Aaij:2014bga}.

This restriction on the usable phase space for TMD factorisation to apply can be avoided by looking at two-particle final states. One example is
the associated  production of a $J/\psi$ or a $\Upsilon$ with a direct photon~\cite{Dunnen:2014eta}. Like for heavy-quark pair and di-jet electroproduction or di-photon hadroproduction, the large scale $Q$ is set by the invariant mass of the system, which can be large
when both the quarkonium and the photon are produced almost back to back with large individual transverse momentum, yet with a small transverse momentum for the pair (its imbalance). This is a rather convenient configuration to be studied experimentally. Moreover, in the case of quarkonium + photon production~\cite{Dunnen:2014eta}, one can enrich the event sample in colour-singlet contributions, which are purely from gluonic interaction, by isolating the quarkonium, since it has a non-zero transverse momentum. This makes it a golden-plated probe to extract gluon TMDs inside unpolarized protons at the LHC. However, along the lines of~\cite{Boer:2014lka}, it may be that colour-octet contributions to quarkonium production associated with a colour singlet particle, like a SM boson ($\gamma$, $W^\pm$, $Z^0$ and $H^0$), could also be treated in the TMD factorisation framework. One of the crucial aspects yet to be fully understood in this case is whether an imbalance to be measured in the final state can be related to the transverse momentum of the initial partons. For this to be true, final-state gluon emissions should admittedly be suppressed at least to a tractable extent.

For unpolarised colliding protons at the LHC, there are two relevant gluon TMDs: the distribution of unpolarised gluons, $f_1^g$, and the distribution of linearly polarised gluons, $h_1^{\perp g}$. The latter is of particular interest as this distribution flips the helicity of the gluon entering the partonic cross section. As a result, the linear polarisation of the gluon manifests itself in two ways: a modification of the transverse-momentum dependence of the cross section in a characteristic way, and an azimuthal modulation of the cross section. As a matter of fact, the linear polarisation may serve as a general new tool in particle physics. Examples of the usefulness of the linear gluon polarisation have been discussed in the context of $H^0$ boson production in~\cite{Boer:2011kf,Boer:2013fca}, as well as for $H^0$+jet production~\cite{Boer:2014lka}.

Like all other TMDs, $f_1^g$ and $h_1^{\perp\,g}$ are affected by the presence of initial and final state interactions, whose effects are encoded in the Wilson lines needed for their gauge-invariant definition. TMD factorisation may therefore fail for some processes, as we already mentioned above. Moreover, TMDs may become process dependent even in those cases where factorisation can be proven. The gluon TMDs appearing in all the proton-proton scattering reactions where only initial state interactions are present, like the ones under study here, correspond to the so called Weisz\"acker-Williams distributions in the small $x$ region. They can be related to gluon TMDs extracted, for example, in heavy quark pair and dijet production in deep-inelastic electron-proton scattering processes~\cite{Boer:2016fqd}. In particular, we expect to probe the same $f_1^g$ and $h_1^{\perp\,g}$ distributions in all such processes, because they are $T$-even TMDs. This is a very important property that still needs to be confirmed by experiments. We refer to~\cite{Boer:2016bfj} for further details.

In this paper, we will explore the relevance of final states consisting of a heavy quarkonium, like $\Upsilon$ or $J/\psi$, produced with a dilepton, be it from a $Z$ boson or from a virtual photon, in the kinematical configuration already mentioned above, such that their transverse momenta are almost back to back. In such a case, we will show how they can help extracting information about the linear polarisation of gluons. Our proposal is motivated by the fact that the detection of a dilepton may experimentally be cleaner or easier compared to the detection of a photon. For instance, the cross section for $J/\psi$ production in association with a $Z$ boson has already been studied by the ATLAS collaboration~\cite{Aad:2014kba} and compared to theoretical predictions~\cite{Lansberg:2016rcx,Gong:2012ah,Mao:2011kf}. This showed that, in the ATLAS acceptance, a significant contribution from double parton scattering (DPS) is expected, which does not fall in the scope of this work. In what follows, we will assume that constraining both observed particles to be back to back makes the DPS yield small enough and we will not venture into the region of large rapidity separations where it can be dominant. Moreover, $J/\psi+\gamma$ and $\Upsilon+\gamma$ have only been investigated~\cite{Aad:2015sda} -- also by ATLAS -- in the context of $H^0$ studies, thus at very large invariant mass where the yields are extremely small and  the contributions we are after are a background to the $H^0$ signal. Studies in the kinematical region considered in~\cite{Dunnen:2014eta} or in~\cite{Li:2008ym,Lansberg:2009db} have not yet been done.
$J/\psi(\Upsilon)+W^\pm$ would also be an option since it has also been experimentally studied -- still by ATLAS~\cite{Aad:2014rua} -- but it is not dominated by gluon induced reactions~\cite{Lansberg:2013wva,Li:2010hc}.

The paper is organised as follows: In Section \ref{analytics} we discuss the gluon fusion process in the TMD factorisation approach for an arbitrary final state and analyse the general structure of the differential cross section. We then analytically calculate the partonic cross sections for each of the azimuthal structures to leading order (LO) accuracy for the final state ${\cal Q}+\ell \bar{\ell}$ where ${\cal Q}$ is a spin-triplet vector quarkonium. In Section \ref{numericsZ}, we give our numerical results for the quarkonium + $Z$ final state, while numerical results for a quarkonium + $(\ell \bar{\ell})$ final state are presented in Section \ref{numericsDY}.  We draw our conclusions in Section~\ref{concl}.

\section{Analytic Calculation within the TMD approach}
\label{analytics}
In this section we consider the process $p(P_a)+p(P_b)\to {\cal Q}(P_{\cal Q})+\ell(l)+\bar{\ell}(\bar{l})+X$, where ${\cal Q}$ denotes a heavy quarkonium bound state, in a kinematical regime where the final state momentum $q^\mu\equiv P_{\cal Q}^\mu+l^\mu+\bar{l}^\mu\equiv P_{\cal Q}^\mu + P_B^\mu$ has a small transverse component, $q_T$, with respect to the beam axis in the proton center-of-mass (c.m.) frame. To be precise, the transverse momentum has to be much smaller than the hard scale of the process, {\it e.g.}\ the invariant mass $Q$ ($q^2=Q^2$) of the final-state, namely $q_T\ll Q$. This is the regime where TMD factorisation can be applied.

Furthermore, we assume that the underlying production mechanism of the heavy quarkonium + lepton pair is due to gluon interactions only. As such, this final state can be considered as a probe for gluon TMDs in proton collisions. In Ref.~\cite{Dunnen:2014eta}, we have shown that, at the LHC,  this is indeed the case for the associated  production  of a heavy quarkonium and a real photon, instead of a lepton pair. 

A second assumption we make here is that the heavy quarkonium is produced directly as a colour-singlet state. For the production of $\Upsilon-\gamma$ at the LHC, this assumption is valid, but not necessarily for $J/\psi-\gamma$~\cite{Dunnen:2014eta}. Where needed, such an assumption can be ensured by isolating the quarkonium (as done in~\cite{Aad:2015sda}). 

\subsection{{General Structure of the Cross Section}}
\label{GeneralCS}
In a first step we present the fully differential cross section in the TMD approach (cf.~Refs.~\cite{Qiu:2011ai,Boer:2011kf,Boer:2013fca,Boer:2010zf,Dunnen:2014eta,Boer:2012bt,Echevarria:2015uaa,Ma:2012hh}) in a general form as
\begin{eqnarray}
\mathrm{d}\sigma_{\mathrm{TMD}}&=&\frac{(2\pi)^4}{S^2}\,\mathrm{dPS}_n\,\frac{1}{(N_c^2-1)^2}\sum_{a,b;I} \mathcal{A}^{ab}_{\mu \nu;I}(\bar{k}_a,\bar{k}_b;\{P_i \})\,\mathcal{A}^{ab\,\ast}_{\rho \sigma;I} (\bar{k}_a,\bar{k}_b;\{P_i\}) \times\nonumber\\
&&\int \frac{d^2 {\bm y}_T}{(2\pi)^2}\,\mathrm{e}^{i{\bm q}_T\cdot {\bm y}_T}\,\tilde{\Gamma}^{\rho\mu}(x_a,{\bm y}_T,\zeta_a,\mu)\,\tilde{\Gamma}^{\sigma\nu}(x_b,{\bm y}_T,\zeta_b,\mu)+\mathcal{O}(q_T/Q)\label{TMDCSyT}\\
&=&\frac{(2\pi)^4}{S^2}\,\mathrm{dPS}_n\, \frac{1}{(N_c^2-1)^2}\sum_{a,b;I} \mathcal{A}^{ab}_{\mu \nu;I}(\bar{k}_a,\bar{k}_b;\{P_i \})\,\mathcal{A}^{ab\,\ast}_{\rho \sigma;I} (\bar{k}_a,\bar{k}_b;\{P_i\}) \times\nonumber\\
&&\hspace*{-2cm}\int \mathrm{d}^2{\bm k}_{aT}\int \mathrm{d}^2 {\bm k}_{bT}\,\delta^{(2)}({\bm k}_{aT}+{\bm k}_{bT}-{\bm q}_T)\,\Gamma^{\rho\mu}(x_a,{\bm k}_{aT},\zeta_a,\mu)\,\Gamma^{\sigma\nu}(x_b,{\bm k}_{bT},\zeta_b,\mu)+\mathcal{O}(q_T/Q)\,, \label{TMDCSkT}
\end{eqnarray}
where $a$ and $b$ are colour indices, $N_c=3$ is the number of colours, and the sum $\sum_I$ denotes the summation over those indices that can identify the particles in the final state, like their helicity.  The differential phase space factor reads $\mathrm{dPS}_n=\prod_{i=1}^n \frac{d^3p_i}{(2\pi)^32E_i}$, with $E_i=\sqrt{{\bm p}_i^2+m_i^2}$.  Moreover, $\mathcal{A}$ denotes the hard scattering amplitude for an arbitrary gluon-induced process $gg\to p_1+...+p_n$, with $n$ colour-singlet particles of momenta $P_1,...,P_n$ in the final state.
It is factorised from the soft part contained in the integral in the second line and can be perturbatively calculated. 
The gluon momenta entering this amplitude are approximated as $\bar{k}_{a/b}^\mu=x_{a/b}P_{a/b}^\mu$. The longitudinal momentum fractions are set to $x_a=q\cdot P_b/P_a\cdot P_b$ and $x_b=q\cdot P_a/P_a\cdot P_b$, with $q = P_1 + P_2 + ... + P_n$.

The formula in (\ref{TMDCSyT}) describes an arbitrary gluon-induced process with a colour-singlet final state in the TMD formalism. Here, the $\tilde{\Gamma}$'s denote the non-perturbative gluonic TMD matrix elements that were properly defined in $y_T$ coordinate space in Ref.~\cite{Echevarria:2015uaa}, including the renormalisation scale $\mu$ as well as an additional scale $\zeta$. The evolution in $\zeta$ of  the TMD correlator $\tilde{\Gamma}$ is governed by the Collins-Soper evolution equation (cf. Refs.~\cite{Collins:2011zzd,Echevarria:2015uaa}). When going from equations (\ref{TMDCSyT}) to (\ref{TMDCSkT}), a simple Fourier transform w.r.t.\ $y_T$ is performed. Hence, $\tilde{\Gamma}$ and $\Gamma$ are related via a Fourier transform. In particular, $\Gamma$ depends on the longitudinal gluon momentum fraction $x$ and the gluon transverse momentum ${\bm k}_T$. 

The TMD correlator $\Gamma$ can be parametrised in terms of gluon TMDs (cf.~Refs.~\cite{Mulders:2000sh,Meissner:2007rx}). For an unpolarised nucleon one finds two structures of the following form,
\begin{equation}
\Gamma^{\mu \nu}(x,{\bm k_T})=\frac{1}{2x}\left( -g_T^{\mu \nu}\ f_1^g(x,{\bm k}_T^2)+\frac{k_T^\mu k_T^\nu+\frac{1}{2}{\bm k}_T^2 g_T^{\mu \nu}}{M^2}h_1^{\perp g}(x,{\bm k}_T^2)\right).\label{Param1}
\end{equation}
The TMD distribution $f_1^g$ can be interpreted as the distribution of unpolarised gluons in an unpolarised nucleon, while the function $h_1^{\perp g}$ is the distribution of linearly polarised gluons \cite{Mulders:2000sh}. In addition, we have introduced the transverse projector $g_T^{\mu \nu}=g^{\mu \nu}-P^{\mu}n^\nu-P^\nu n^\mu$, where $P$  is the nucleon momentum and $n$ is an adjoint light-cone vector such that $n^2=0$ and $P\cdot n=1$.  Moreover, in (\ref{Param1}), $M$ denotes the nucleon mass. We note that the parametrisation (\ref{Param1}) is slightly different to the one in Ref.~\cite{Echevarria:2015uaa} where the $M^2$ in (\ref{Param1}) is replaced by $\frac{1}{2}{\bm{k}_T^2}$. Also, we have an additional factor $1/x$ in (\ref{Param1}). 

Getting back to the expression for the cross section in (\ref{TMDCSkT}), it is useful to consider it in the c.m.~frame of both incoming protons, with $P_{a/b}$ along the positive (negative) $z$-direction, e.g., $P^\mu_{a/b}=\sqrt{S/2}n^\mu_{\pm}$, with $n^{\mu}=(1,0,0,\pm 1)/\sqrt{2}$. In this frame, we can work with simple polarisation vectors of the gluons (which are approximated to be collinear to the proton momenta in the amplitudes $\mathcal{A}$ in (\ref{TMDCSkT})). One can easily find that these polarisation vectors acquire the following form in terms of the gluon helicities $\lambda_{a/b}=\pm 1$,
\begin{eqnarray}
\varepsilon_{\lambda_a}^\mu(\bar{k}_a)=\left(0,-\frac{\lambda_a}{\sqrt{2}},-\frac{i}{\sqrt{2}},0\right)&;&\varepsilon_{\lambda_b}^\mu(\bar{k}_b)=\left(0,\frac{\lambda_b}{\sqrt{2}},-\frac{i}{\sqrt{2}},0\right).\label{PolVec}
\end{eqnarray}
These polarisation vectors lead to the common polarisation sum in the Feynman gauge, 
\begin{equation}
-g_T^{\mu \nu}=\sum_{\lambda_{a/b}}\varepsilon^\mu_{\lambda_{a/b}}(\bar{k}_{a/b})\,  \varepsilon^{\nu\,\ast}_{\lambda_{a/b}}(\bar{k}_{a/b}).\label{PolSum}
\end{equation}
In addition, the polarisation vectors are transverse, i.e., $\bar{k}_a\cdot \varepsilon_{\lambda_a}(\bar{k}_a)=\bar{k}_b\cdot \varepsilon_{\lambda_a}(\bar{k}_a)=\bar{k}_a\cdot \varepsilon_{\lambda_b}(\bar{k}_b)=\bar{k}_b\cdot \varepsilon_{\lambda_b}(\bar{k}_b)=0$. Of course, the polarisation vectors (\ref{PolVec}) are only determined up to a phase, therefore other realisations are possible as well.

Since the contraction of Minkowski indices $\mu, \nu, \rho, \sigma$ in (\ref{TMDCSkT}) is to be understood as a contraction in the {\it transverse} space, we can insert the polarisation sums (\ref{PolSum}) and rewrite (\ref{TMDCSkT}) in terms of the helicity amplitudes,
\begin{eqnarray}
&&\mathrm{d}\sigma_{\mathrm{TMD}}=\frac{(2\pi)^4}{S^2}\,\mathrm{dPS}_n\!\!\!\sum_{\lambda_a,\bar{\lambda}_a,\lambda_b,\bar{\lambda}_b=\pm 1}\,\frac{1}{(N_c^2-1)^2}\sum_{a,b;I} \mathcal{A}^{ab}_{\lambda_a, \lambda_b;I}(\bar{k}_a,\bar{k}_b;\{P_i \})\,\mathcal{A}^{ab\,\ast}_{\bar{\lambda}_a, \bar{\lambda}_b;I}(\bar{k}_a,\bar{k}_b;\{P_i\}) \times\label{TMDCSPolVec}\\
&&\int \mathrm{d}^2{\bm k}_{aT}\int \mathrm{d}^2 {\bm k}_{bT}\,\delta^{(2)}({\bm k}_{aT}+{\bm k}_{bT}-{\bm q}_T)\,\Gamma_{\bar{\lambda}_a,\lambda_a}(x_a,{\bm k}_{aT},\zeta_a,\mu)\,\Gamma_{\bar{\lambda}_b,\lambda_b}(x_b,{\bm k}_{bT},\zeta_b,\mu)+\mathcal{O}(q_T/Q).~~~~~~\nonumber
\end{eqnarray}
Here, we introduced the notation $\mathcal{A}_{\lambda_a,\lambda_b}\equiv \varepsilon^\mu_{\lambda_a}(\bar{k}_a)\ \varepsilon^\nu_{\lambda_b}(\bar{k}_b)\ \mathcal{A}_{\mu \nu}$ for the helicity amplitudes, and the \emph{helicity} gluon correlator $\Gamma_{\bar{\lambda},\lambda}\equiv (\varepsilon^\mu_{\bar{\lambda}})(\bar{k})\, \varepsilon^{\nu\,\ast}_\lambda (\bar {k})\,\Gamma_{\mu \nu}$. For an unpolarised nucleon the parameterisation in terms of the gluon helicities then takes the following form,
\begin{eqnarray}
\Gamma_{\bar{\lambda}_a,\lambda_a}(x_a,{\bm k}_{aT})&=&\frac{1}{2x_a}\left (\delta_{\lambda_a,\bar{\lambda}_a}\ f_1^g(x_a,{\bm k}_{aT}^2)+\frac{k_{ax}^2-k_{ay}^2-2i\lambda_a k_{ax}k_{ay}}{2M^2}\delta_{\lambda_a,-\bar{\lambda}_a}h_1^{\perp g}(x_a,{\bm k}_{aT}^2)\right),\label{Param2a}\\
\Gamma_{\bar{\lambda}_b,\lambda_b}(x_b,{\bm k}_{bT})&=&\frac{1}{2x_b}\left( \delta_{\lambda_b,\bar{\lambda}_b}\ f_1^g(x_b,{\bm k}_{bT}^2)+\frac{k_{bx}^2-k_{by}^2+2i\lambda_b k_{bx}k_{by}}{2M^2}\delta_{\lambda_b,-\bar{\lambda}_b}h_1^{\perp g}(x_b,{\bm k}_{bT}^2)\right ),\label{Param2b}
\end{eqnarray}
with ${\bm k}_{aT}=(k_{ax},k_{ay})$ and ${\bm k}_{bT}=(k_{bx},k_{by})$. It is evident from (\ref{Param2a},\ref{Param2b}) that the gluon TMD $f_1^g$ conserves the gluon helicity at the non-perturbative level, while the the distribution of linearly polarised gluons, $h_1^{\perp g}$, flips it. We can insert the parameterisations (\ref{Param2a},\ref{Param2b}) into (\ref{TMDCSPolVec}) and write the differential cross section in the general form
\begin{eqnarray}
\mathrm{d}\sigma_{\mathrm{TMD}}&=&\frac{(2\pi)^4}{4x_a x_b S^2}\,\mathrm{dPS}_n \Big \{\hat{F}_1(\bar{k}_a,\bar{k}_b;\{P_i\})\,\mathcal{C}[f_1^g\,f_1^g]+ \hat{F}_2(\bar{k}_a,\bar{k}_b;\{P_i\})\,\mathcal{C}[w_2\,h_1^{\perp g}\,h_1^{\perp g}]\nonumber\\
&&+\hat{F}_{3a}(\bar{k}_a,\bar{k}_b;\{P_i\})\,\mathcal{C}[w_{3a} \,h_1^{\perp g}\,f_1^g]+\hat{F}_{3b}(\bar{k}_a,\bar{k}_b;\{P_i\})\,\mathcal{C}[w_{3b} \,f_1^g\,h_1^{\perp g}]\nonumber\\
&&+\hat{F}_{4}(\bar{k}_a,\bar{k}_b;\{P_i\})\,\mathcal{C}[w_{4} \,h_1^{\perp g}\,h_1^{\perp g}] \Big \}+\mathcal{O}(q_T/Q)\,,\label{Structures}
\end{eqnarray}
where the different transverse momentum convolutions of gluon TMDs are abbreviated as
\begin{equation}
\mathcal{C}[w\,f\,g]\equiv \int \mathrm{d}^2{\bm k}_{aT}\int \mathrm{d}^2{\bm k}_{bT}\,\delta^{(2)}({\bm k}_{aT}+{\bm k}_{bT}-{\bm q}_T)\,w({\bm k}_{aT},{\bm k}_{bT},{\bm q}_{T})\,f(x_a,{\bm k}_{aT}^2)\,g(x_b,{\bm k}_{bT}^2)\,.\label{Conv}
\end{equation}
Since the distribution of linearly polarised gluons carries gluon transverse-momentum-dependent prefactors in the parameterisations (\ref{Param2a},\ref{Param2b}), these prefactors emerge again in the convolutions in (\ref{Structures}) in the weighting factors $w$. To be specific we have 
\begin{eqnarray}
w_2=\frac{2({\bm k}_{aT}\cdot {\bm k}_{bT})^2-{\bm k}_{aT}^2\,{\bm k}_{bT}^2}{4M^4}&,& w_4=  2\left[\frac{\bm k_{aT} \cdot \bm k_{bT}}{2M^2} - \frac{ (\bm k_{aT} \cdot \bm q_T) (\bm k_{b T} \cdot \bm q_T)} {M^2 \bm{q}_T^2}\right]^2 -\frac{\bm k_{aT}^2\bm k_{bT}^2 }{4 M^4}
		\,, \nonumber\\
w_{3a}=\frac{{\bm k}_{aT}^2 {\bm q}_T^2-2({\bm q}_T\cdot {\bm k}_{aT})^2}{2M^2\,{\bm q}_T^2}&,&w_{3b}=\frac{{\bm k}_{bT}^2 {\bm q}_T^2-2({\bm q}_T\cdot {\bm k}_{bT})^2}{2M^2\,{\bm q}_T^2}.\label{weights}
\end{eqnarray}
We find five different structures in the cross section in (\ref{Structures}): the first one is given by a convolution of two unpolarised gluon TMDs, while the second and fifth are given by the convolutions of two distributions of linearly polarised gluons. In the latter case the helicities of each gluon are flipped. We separated both these structures because the second typically provides azimuthally isotropic contributions, while the fifth leads to azimuthal modulations (around the beam axis) of the differential cross section. The third and fourth structures are mixed convolutions of an unpolarised and a linearly polarised gluon TMD. Here, only one of the gluon helicities is flipped.

The factors $\hat{F}_i$ in (\ref{Structures}) can be calculated perturbatively since they are defined in terms of the partonic amplitudes $\mathcal{A}$ in the following way,
\begin{eqnarray}
\hat{F}_1(\bar{k}_a,\bar{k}_b;\{P_i\}) &=&\sum_{\lambda_a,\lambda_b,=\pm 1} \frac{1}{(N_c^2-1)^2}\sum_I \mathcal{A}^{ab}_{\lambda_a, \lambda_b;I}(\bar{k}_a,\bar{k}_b;\{P_i \})\,\mathcal{A}^{ab\,\ast}_{\lambda_a,\lambda_b;I}(\bar{k}_a,\bar{k}_b;\{P_i\}) \,,\nonumber\\
\hat{F}_2(\bar{k}_a,\bar{k}_b;\{P_i\}) &=&\sum_{\lambda=\pm 1} \frac{1}{(N_c^2-1)^2}\sum_I \mathcal{A}^{ab}_{\lambda, \lambda;I}(\bar{k}_a,\bar{k}_b;\{P_i \})\,\mathcal{A}^{ab\,\ast}_{-\lambda,-\lambda;I}(\bar{k}_a,\bar{k}_b;\{P_i\}) \,,\nonumber\\
\hat{F}_{3a}(\bar{k}_a,\bar{k}_b;\{P_i\}) &=&\sum_{\lambda_a,\lambda_b,=\pm 1} \frac{1}{(N_c^2-1)^2}\sum_I \mathcal{A}^{ab}_{\lambda_a, \lambda_b;I}(\bar{k}_a,\bar{k}_b;\{P_i \})\,\mathcal{A}^{ab\,\ast}_{-\lambda_a,\lambda_b;I}(\bar{k}_a,\bar{k}_b;\{P_i\}) \,,\nonumber\\
\hat{F}_{3b}(\bar{k}_a,\bar{k}_b;\{P_i\}) &=&\sum_{\lambda_a,\lambda_b,=\pm 1}\frac{1}{(N_c^2-1)^2}\sum_I \mathcal{A}^{ab}_{\lambda_a, \lambda_b;I}(\bar{k}_a,\bar{k}_b;\{P_i \})\,\mathcal{A}^{ab\,\ast}_{\lambda_a,-\lambda_b;I}(\bar{k}_a,\bar{k}_b;\{P_i\}) \,,\nonumber\\
\hat{F}_4(\bar{k}_a,\bar{k}_b;\{P_i\}) &=&\sum_{\lambda=\pm 1} \frac{1}{(N_c^2-1)^2}\sum_I \mathcal{A}^{ab}_{\lambda, -\lambda;I}(\bar{k}_a,\bar{k}_b;\{P_i \})\,\mathcal{A}^{ab\,\ast}_{-\lambda,\lambda;I}(\bar{k}_a,\bar{k}_b;\{P_i\}) \,.\label{DefF}
\end{eqnarray}

The results in (\ref{Structures}) - (\ref{DefF}) are the main ones of this subsection, and are valid for an arbitrary process induced by the interaction of two gluons in the initial state, where a colour-singlet final state is produced. In the following, we will analyse a specific hadronic reaction in which a heavy quarkonium state and a lepton pair are produced back to back in transverse space. 

\subsection{The subprocess $gg\to {\cal Q}\,\ell \,\bar{\ell}$}

\subsubsection{The helicity structure}

We will now compute the amplitude $\mathcal{A}$ for the process $gg\to {\cal Q}\,\ell\, \bar{\ell}$ and use these results to determine the explicit expressions of the $\hat{F}_i$ in (\ref{Structures}).  The first observation we make is that the lepton pair is produced either through the decay of a virtual photon or a $Z$ boson. Hence, we can decompose the amplitude further and write
\begin{equation}
\mathcal{A}^{gg\to {\cal Q}\,\ell\, \bar{\ell}}_{\lambda_a,\lambda_b}=\sum_{j=\gamma^\ast,Z}i\,C_j\,\mathcal{A}^{gg\to {\cal Q}\,j;\,\mu}_{\lambda_a,\lambda_b}\,\frac{g_{\mu \nu}-P_{B\mu}P_{B\nu}/M_B^2}{M_B^2-\Delta_j}\,\bar{u}(l,\sigma)\gamma^{\nu}(a_j+b_j\gamma_5)v(\bar{l},\bar{\sigma}).\label{AmpZ}
\end{equation}

With the decomposition (\ref{AmpZ}) at hand we can calculate the amplitude squared,
\begin{eqnarray}
\mathcal{F}_{\lambda_a\lambda_b;\bar{\lambda}_a\bar{\lambda}_b}&\equiv&\frac{(2\pi)^4}{4x_a x_b S^2}\,\frac{\mathrm{dPS}_3}{(N_c^2-1)^2}\sum_I \mathcal{A}^{gg\to {\cal Q}\,\ell \,\bar{\ell}}_{\lambda_a, \lambda_b;I}\,\mathcal{A}^{\ast \,gg\to {\cal Q}\,\ell \,\bar{\ell}}_{\bar{\lambda}_a,\bar{\lambda_b};I} \nonumber\\
&=&\frac{1}{4(2\pi)^5x_a x_b S^2(N_c^2-1)^2}\,\frac{\mathrm{d}^3P_{\cal Q}}{2E_Q}\times\nonumber\\
&&\Bigg(\sum_{j,\bar{j}=\gamma^\ast,Z}C_jC_{\bar{j}}\,\sum_I \mathcal{A}^{gg\to {\cal Q}\,j;\,\mu}_{\lambda_a,\lambda_b;I}\mathcal{A}^{\ast \,gg\to {\cal Q}\,\bar{j};\,\rho}_{\bar{\lambda}_a,\bar{\lambda}_b;I}\; \frac{g_{\mu \nu}-P_{B\mu}P_{B\nu}/M_B^2}{(M_B^2-\Delta_j)}\; \frac{g_{\rho \sigma}-P_{B\rho}P_{B\sigma}/M_B^2}{(M_B^2-\Delta_{\bar{j}})^\ast}\times\nonumber\\
&&\left[L_{j\bar{j}}^{\nu \sigma}(l,\bar{l})\,\mathrm{d}^4l\,\delta(l^2)\,\theta(l^0)\,\mathrm{d}^4\bar{l}\,\delta(\bar{l}^2)\theta(\bar{l}^0)\right]\Bigg).\label{auxF}
\end{eqnarray}
We note that any explicit dependence on the single lepton momenta is hidden in the last line, where we have introduced the Lorentz-invariant leptonic tensor 
\begin{equation}
L_{j\bar{j}}^{\nu \sigma}=4(a_j a_{\bar{j}}+b_j b_{\bar{j}})\,\left(l^\nu \bar{l}^\sigma + l^\sigma \bar{l}^\nu-\tfrac{1}{2}M_B^2g^{\nu\sigma}\right)+4(a_j b_{\bar{j}}+b_j a_{\bar{j}})\,\epsilon^{\nu\sigma \eta \kappa}l_\eta\bar{l}_\kappa\,.\label{LeptTens}
\end{equation}
We now integrate over the momentum $\bar{l}$ of the antilepton, keeping in mind that $\mathrm{d}^4l\,\mathrm{d}^4\bar{l}\,\delta(l^2)\,\delta(\bar{l}^2)=\mathrm{d}^4P_B\,\mathrm{d}^4\bar{l}\,\delta(\bar{l}^2)\,\delta((P_B-\bar{l})^2)$. Since the last line of (\ref{auxF}) is Lorentz-invariant, this is conveniently done  in the center-of-mass frame of the lepton pair where the pair momentum takes the simple form $P_B^\mu=(M_B,0,0,0)$. The delta functions are easily integrated out, and we find $\mathrm{d}^4l\,\mathrm{d}^4\bar{l}\,\delta(l^2)\,\delta(\bar{l}^2)\to \frac{1}{8}\,\mathrm{d}^4P_B\,\mathrm{d}\Omega$, with $\Omega$ the solid angle which determines the direction of the antilepton. Using an explicit form of the antilepton momentum $\bar{l}^\mu=\frac{M_B}{2}(1,\vec{e})$ with $\vec{e}=(\sin\theta\, \cos\phi,\sin\theta\,\sin\phi,\cos\theta)$ in this specific dilepton c.m.-frame we can perform the solid angle integration and obtain
\begin{eqnarray}
\int \mathrm{d}\Omega\,L^{\nu \sigma}_{j\bar{j}}&=&\int \mathrm{d}\Omega\,4\left [ (a_j a_{\bar{j}}+b_j b_{\bar{j}})\,\left(-2\bar{l}^\nu \bar{l}^\sigma +P_B^\nu\bar{l}^\sigma+ P_B^\sigma \bar{l}^\nu-\tfrac{1}{2}M_B^2g^{\nu\sigma}\right)+(a_j b_{\bar{j}}+b_j a_{\bar{j}})\,\epsilon^{\nu\sigma \eta \kappa}P_{B\eta}\bar{l}_\kappa \right ]\nonumber\\
&=&\frac{16}{3}\pi (a_j a_{\bar{j}}+b_j b_{\bar{j}})\,\left(P_B^\nu P_B^\sigma-M_B^2g^{\nu \sigma}\right).\label{IntSolidAngle}
\end{eqnarray}
We insert this result into the quantity $\mathcal{F}$ in (\ref{auxF}) and perform the contraction with the numerators of the weak boson-propagators. This leads to a factor $-M_B^2(g_{\mu\rho}-P_{B\mu}P_{B\rho}/M_B^2)$ coming from $M_B^2\sum_{\lambda_B=0,\pm 1}\varepsilon_{\mu,\lambda_B}^\ast(P_B)\,\varepsilon_{\rho,\lambda_B}(P_B)$ in the numerator, which we have indentified as the  sum over the helicities of the polarisation vectors of the virtual electroweak boson. Note that, since the virtual electroweak boson is massive (formally carrying a mass $P_B^2=M_B^2$), three polarisations are necessary. Utilising the polarisation sum in this way allows us to work with helicity amplitudes when calculating the process $gg\to {\cal Q}\,\gamma^\ast / Z$. 

Finally, we also need to modify the remaining phase space $\mathrm{d}^4P_{\cal Q}\,\mathrm{d}^4P_B\,\delta(P_{\cal Q}^2-M_{\cal Q}^2)\theta(P_{\cal Q}^0)$ where $M_{\cal Q}$ is the mass of the heavy quarkonium state. We do so by considering the relative momenta $q^\mu=P_{\cal Q}^\mu+P_B^\mu$ and $\Delta q^\mu=P_{\cal Q}^\mu-P_B^\mu$ and note that $\mathrm{d}^4P_B\,\mathrm{d}^4P_{\cal Q}=\frac{1}{16}\mathrm{d}^4q\,\mathrm{d}^4\Delta q$. It is most convenient to analyse these factors in a c.m.-frame of the heavy quarkonium and the virtual electroweak gauge boson, such as the Collins-Soper (CS) frame. In this frame the gluon momenta are along the $z$-axis whereas the heavy quarkonium and electroweak gauge boson momenta have an explicit representation in this frame, $P_{\cal Q}^\mu=\left(\sqrt{\Lambda^2+M_{\cal Q}^2},\Lambda\,\vec{e}\right)$ and  $P_B^\mu=\left(\sqrt{\Lambda^2+M_B^2},-\Lambda\,\vec{e}\right)$, with $\vec{e}=(\sin\theta\, \cos\phi,\sin\theta\,\sin\phi,\cos\theta)$ as before, where $\theta$ and $\phi$ are the Collins-Soper angles. From considering $q^2=(P_{\cal Q}+P_B)^2\equiv Q^2>0$ we find that $\Lambda=\sqrt{\lambda(Q^2,M_{\cal Q}^2,M_B^2)}/(2Q)$, with $\lambda(x,y,z)=x^2+y^2+z^2-2xy-2xz-2yz$. Hence, the relative momenta take the explicit form $q^\mu=(Q,0,0,0)$ and $\Delta q^\mu=((M_{\cal Q}^2-M_B^2)/Q,2\Lambda \vec{e})$ in the CS frame, and we conclude that $\mathrm{d}^4P_{\cal Q}\,\mathrm{d}^4P_B\,\delta(P_{\cal Q}^2-M_{\cal Q}^2)\theta(P_{\cal Q}^0)\to \Lambda/(4Q)\,\mathrm{d}^4q\,\mathrm{d}M_B^2\,\mathrm{d}\Omega$.
Collecting all the  results above, we rewrite (\ref{auxF}) as
\begin{eqnarray}
\mathcal{F}_{\lambda_a\lambda_b;\bar{\lambda}_a\bar{\lambda}_b}
&=&\frac{\Lambda\,M_B^2}{48\,(2\pi)^4\,x_a x_b S^2\,Q\,(N_c^2-1)^2}\,\mathrm{d}^4q\,\mathrm{d}M_B^2\,\mathrm{d}\Omega\times\nonumber\\
&&\Bigg(\sum_{j,\bar{j}=\gamma^\ast,Z}\frac{C_jC_{\bar{j}}\,(a_j a_{\bar{j}}+b_j b_{\bar{j}})}{(M_B^2-\Delta_j)\,(M_B^2-\Delta_{\bar{j}})^\ast}\,\sum_{\lambda_B=0,\pm1;I} \mathcal{A}^{gg\to {\cal Q}\,j}_{\lambda_a,\lambda_b;\lambda_B;I} \mathcal{A}^{\ast\,gg\to {\cal Q}\,\bar{j}}_{\bar{\lambda}_a,\bar{\lambda}_b;\lambda_B;I} \Bigg).\label{auxF1}
\end{eqnarray}

\begin{figure}[t]
\centering
\includegraphics[width=1\textwidth,angle=0]{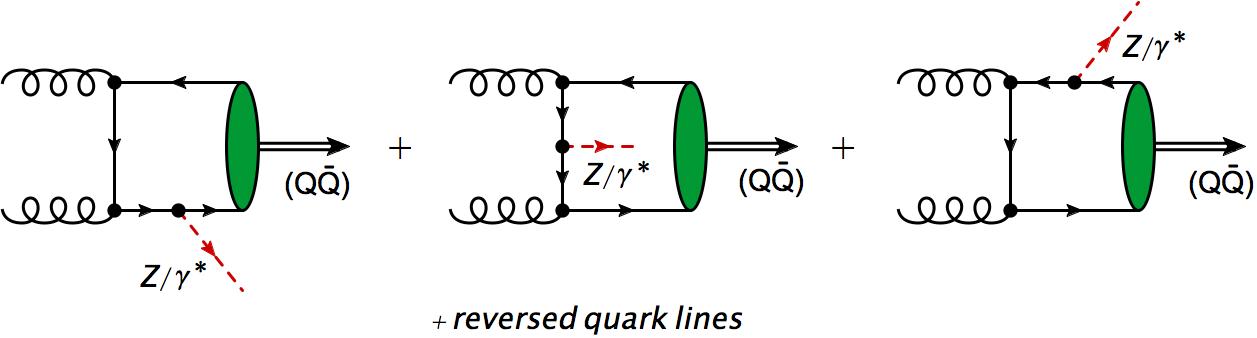}
\caption{Leading order diagrams for the subprocess $gg\to {\cal Q}\,Z/\gamma^\ast$. Diagrams where the direction of the quark lines are reversed also contribute.\label{fig:JPsiZ}}
\end{figure}

\subsubsection{The amplitude $\mathcal{A}^{gg\to {\cal Q}\,(Z/\gamma^\ast)}$}

As a final step, we need to calculate the amplitude $\mathcal{A}^{gg\to {\cal Q}\,(Z/\gamma^\ast)}$. We will do so to leading order accuracy in perturbative QCD and assume a colour-singlet heavy quarkonium state. The leading order diagrams are shown in Fig.~\ref{fig:JPsiZ}. In the following we restrict ourselves to heavy quarkonium states without orbital angular momentum quantum numbers, in particular to a $J/\psi$ or $\Upsilon$ state. 
In the colour-singlet model one typically neglects relative momenta of the heavy quark-antiquark pair in the hard part such that the wave function of the heavy quarkonium state shrinks to the origin, and the vertex of the transition of both heavy quarks forming a $J/\psi$ or $\Upsilon$ $\--$ the green blobs in Fig.~\ref{fig:JPsiZ} $\--$ reduces to a vertex $1/\sqrt{4\pi N_c M_{\cal Q}}\,R_0(0)\,(\kks{P}_{\cal Q}-M_{\cal Q})\,\kks{\varepsilon}_{\lambda_{\cal Q}}(P_{\cal Q})$, where $R_0(0)$ is the radial wave function of the heavy quarkonium state at the origin in the quarkonium rest frame, and $\varepsilon_{\lambda_{\cal Q}}(P_{\cal Q})$ the polarisation vector of the $J/\psi$ or $\Upsilon$ (cf.~Ref.~\cite{Boer:2012bt}). Note that we have again three polarisations $\lambda_{\cal Q}=0,\pm 1$ for a massive spin-1 particle. We also note that the coupling of the electroweak gauge boson depends on the flavor of the quark, i.e., we have a quark-gauge boson vertex of the form $\gamma^\mu (a_q+b_q \gamma_5)$, with
$a_c=-5/3+(8/3)m_W^2/m_Z^2$, $b_c=-1$ for a charm quark coupling to a $Z$-boson, $a_b=7/3-(4/3)m_W^2/m_Z^2$, $b_b=+1$ for a bottom quark coupling to a $Z$-boson, and $a_c=a_b=1$, $b_c=b_b=0$ for a quark coupling to a photon.

The leading order diagrams can then be calculated in a straightforward way, and we will not elaborate on the details of the calculation. Once we have obtained the analytic expressions for the helicity amplitudes $\mathcal{A}^{gg\to J/\psi[\Upsilon]\,(Z/\gamma^\ast)}$ we insert them into (\ref{auxF1}). It is then only a minor step to extract the $\hat{F}_i$-prefactors in (\ref{DefF}). 

\subsubsection{The azimuthal dependence of the differential cross section}

To go further, we decompose, in the CS frame, the differential cross section in terms of factors $\hat{F}_i$ :
\begin{eqnarray}
\frac{\mathrm{d}\sigma^{pp\to J/\psi[\Upsilon]\ell \bar{\ell}X}_{\mathrm{TMD,\,LO}}}{\mathrm{d}^4q\,\mathrm{d}M_B^2\,\mathrm{d}\Omega}&=&\hat{F}_1(Q,\alpha,\beta,\theta)\,\mathcal{C}[f_1^g\,f_1^g]+ \hat{F}_2(Q,\alpha,\beta,\theta)\,\mathcal{C}[w_2\,h_1^{\perp g}\,h_1^{\perp g}]\nonumber\\
&&+\left \{ \hat{F}_{3a}(Q,\alpha,\beta,\theta)\,\mathcal{C}[w_{3a} \,h_1^{\perp g}\,f_1^g]+\hat{F}_{3b}(Q,\alpha,\beta,\theta)\,\mathcal{C}[w_{3b} \,f_1^g\,h_1^{\perp g}]\right \}\,\cos 2\phi \nonumber\\
&&+\hat{F}_{4}(Q,\alpha,\beta,\theta)\,\mathcal{C}[w_{4} \,h_1^{\perp g}\,h_1^{\perp g}]\,\cos 4\phi,\label{StructuresPro}
\end{eqnarray}
where the $\hat{F}_i$ are functions of the $\theta$ CS-angle, the parameters $\alpha\equiv M_{\cal Q}/Q$ and $\beta=M_B/Q$. The
prefactors  $\hat{F}_i$ have the form 
\begin{equation}
\hat{F}_i=\frac{4\alpha_s^2\,|R_0(0)|^2\,\Lambda\,M_B^2}{3\,\pi^3\,x_a x_b S^2\,Q^3\,M_{\cal Q}\,N_c\,(N_c^2-1)^2}\sum_{j,\bar{j}=\gamma^\ast,Z}\frac{\tilde{C}_j\tilde{C}_{\bar{j}}\,(a_j a_{\bar{j}}+b_j b_{\bar{j}})}{(M_B^2-\Delta_j)\,(M_B^2-\Delta_{\bar{j}})^\ast}\,\frac{\hat{f}_i(\alpha,\beta,\theta)}{\hat{D}(\alpha,\beta,\theta)},\label{Fis}
\end{equation}
where $\tilde{C}_{\gamma^\ast}=-4\pi \alpha_{\mathrm{em}}e_q$ and $\tilde{C}_Z=m_Z^2 G_F a_q/(2\sqrt{2})$, and $\Lambda=\frac{Q}{2}\sqrt{\lambda(1,\alpha^2,\beta^2)}$. The functions $\hat{f}_i$, $\hat{D}$ acquire the following form in terms of the auxiliary variables $A\equiv 1+\alpha^2-\beta^2$ and $B\equiv\sqrt{A^2-4\alpha^2}=\sqrt{\lambda(1,\alpha^2,\beta^2)}\ge 0$,
\begin{eqnarray}
\hat{D}(\alpha,\beta,\theta)&=&(2-A)^2\,\left(A^2-B^2\,\cos^2 \theta \right)^2\,,\label{D}
\end{eqnarray}
as well as
\begin{eqnarray}
\hat{f}_1(\alpha,\beta,\theta)&=&a_{1,0}+a_{1,2}\,\sin^2 \theta +a_{1,4}\,\sin^4 \theta,\nonumber\\
\hat{f}_2(\alpha,\beta,\theta)&=&a_{2,0},\nonumber\\
\hat{f}_{3a}(\alpha,\beta,\theta)&=&a_{3,0} \sin
   ^2 \theta \,=\,\hat{f}_{3b}(\alpha,\beta,\theta),\nonumber\\
\hat{f}_4(\alpha,\beta,\theta) &=& a_{4,0} \sin ^4 \theta ,\label{fs}
\end{eqnarray}
with
\begin{eqnarray}
a_{1,0}&=&16\alpha^2 \left [ 2(1-\alpha^2)(1-\alpha^2+\beta^2)+\alpha^2\beta^2\right ],\nonumber\\
a_{1,2}&=&8\left [ (\alpha-\beta)^2-1\right ]\,\left [(\alpha+\beta)^2-1\right ]\,\left [ (\alpha^2-\beta^2)^2-2\alpha^2\right ],\nonumber\\
a_{1,4}&=& \left(2 \alpha ^2+\beta ^2\right) \left [1-2(\alpha^2+\beta^2)+(\alpha^2-\beta^2)^2\right ]^2,\nonumber\\
a_{2,0}&=&48\alpha^4\beta^2,\nonumber\\
a_{3,0}&=&8\alpha^2(\alpha^2+2\beta^2)  \left [1-2(\alpha^2+\beta^2)+(\alpha^2-\beta^2)^2\right ],\nonumber\\
a_{4,0}&=&a_{1,4}.\label{as}
\end{eqnarray}

Note that in the limit of real photon production, that is, $M_B\to 0$ or $\beta\to 0$, we recover the results of the $\hat{F_i}$ that were found in Ref.~\cite{Dunnen:2014eta}. We find that the partonic prefactor $\hat{f}_2$ vanishes in the real photon limit $\--$ a unique feature of this particular final state.

Eqs.~(\ref{StructuresPro}-\ref{as}) constitute the main analytical results of this work. It is however instructive to analyse the cross section that is integrated over the CS angles including a possible azimuthal weighting factor. These weighting factors may enable us to disentangle the various azimuthal contributions in (\ref{StructuresPro}). For example, analoguously to Ref.~\cite{Dunnen:2014eta}, we can deduce the following weighted cross sections from (\ref{StructuresPro}),
\begin{eqnarray}
N^{(0)}\equiv \int \mathrm{d}\Omega \frac{\mathrm{d}\sigma^{pp\to J/\psi[\Upsilon]\ell \bar{\ell}X}_{\mathrm{TMD,\,LO}}}{\mathrm{d}^4q\,\mathrm{d}M_B^2\,\mathrm{d}\Omega} = \frac{\mathrm{d}\sigma^{pp\to J/\psi[\Upsilon]\ell \bar{\ell}X}_{\mathrm{TMD,\,LO}}}{\mathrm{d}^4q\,\mathrm{d}M_B^2}\!\!&=&\!\! \hat{F}_1(Q,\alpha,\beta)\,\mathcal{C}[f_1^g\,f_1^g]+\hat{F}_2(Q,\alpha,\beta)\,\mathcal{C}[w_2\,h_1^{\perp g}\,h_1^{\perp g}]\,,\nonumber\\
N^{(2)}\equiv \int \mathrm{d}\Omega \cos 2\phi \,\frac{\mathrm{d}\sigma^{pp\to J/\psi[\Upsilon]\ell \bar{\ell}X}_{\mathrm{TMD,\,LO}}}{\mathrm{d}^4q\,\mathrm{d}M_B^2\,\mathrm{d}\Omega}\!\!&=&\!\!\hat{F}_3(Q,\alpha,\beta)\,\left(\mathcal{C}[w_{3a}\,h_1^{\perp g}\,f_1^g]+\mathcal{C}[w_{3b}\,f_1^g\,h_1^{\perp g}]\right)\,,\nonumber\\
N^{(4)}\equiv \int \mathrm{d}\Omega \cos 4\phi \,\frac{\mathrm{d}\sigma^{pp\to J/\psi[\Upsilon]\ell \bar{\ell}X}_{\mathrm{TMD,\,LO}}}{\mathrm{d}^4q\,\mathrm{d}M_B^2\,\mathrm{d}\Omega}\!\!&=&\!\!\hat{F}_4(Q,\alpha,\beta)\,\mathcal{C}[w_{4}\,h_1^{\perp g}\,h_1^{\perp g}]\,.\label{WeightedStructures}
\end{eqnarray}
Note that we use the same symbols $\hat{F}_i$ for the integrated or weighted partonic prefactors, i.e., $\hat{F}_{1,2}(Q,\alpha,\beta)=2\pi \int \mathrm{d}\theta\,\hat{F}_{1,2}(Q,\alpha,\beta,\theta)$ and $\hat{F}_{3,4}(Q,\alpha,\beta)=\pi \int \mathrm{d}\theta\,\hat{F}_{3a/b,4}(Q,\alpha,\beta,\theta)$. The $\theta$-integration can be performed analytically, and we find the following integrated partonic prefactors utilising Eqs.~(\ref{Fis}-\ref{as}),
\begin{eqnarray}
\hat{F}_1(Q,\alpha,\beta)&=&\frac{\hat{F}}{(2-A)^2A^2B^4}\left [\frac{B^4\,a_{1,0}}{4\alpha^2}-B^2\,a_{1,2}+(3A^2-B^2)\,a_{1,4} \right .\nonumber\\
&&+ \left . \left(B^4\, a_{1,0}+B^2(A^2+B^2)\,a_{1,2}-4\alpha^2\,(3A^2+B^2)\,a_{1,4}\right)\,\frac{\ln\left(\tfrac{A+B}{A-B}\right)}{2AB}\right ]\,,\label{F1int}
\end{eqnarray}
\begin{eqnarray}
\hat{F}_2(Q,\alpha,\beta)&=&\frac{\hat{F}\,a_{2,0}}{(2-A)^2A^2}\left [\frac{1}{4\alpha^2}+\frac{\ln\left(\tfrac{A+B}{A-B}\right)}{2AB}\right ]\,,\label{F2int}\\
\hat{F}_3(Q,\alpha,\beta)&=&-\frac{\hat{F}\,a_{3,0}}{2(2-A)^2A^2B^2}\left[1-(A^2+B^2)\frac{\ln\left(\tfrac{A+B}{A-B}\right)}{2AB}\right ]\,,\label{F3int}\\
\hat{F}_4(Q,\alpha,\beta)&=&\frac{\hat{F}\,a_{4,0}}{2(2-A)^2A^2B^4}\left [3A^2-B^2-4\alpha^2\,(3A^2+B^2)\frac{\ln\left(\tfrac{A+B}{A-B}\right)}{2AB}\right ]\,.\label{F4int}
\end{eqnarray}
where we have used the definition
\begin{equation}
\hat{F} \equiv \frac{8\alpha_s^2\,|R_0(0)|^2\,\Lambda\,M_B^2}{3\,\pi^2\, x_a x_b S^2\,Q^3\,M_{\cal Q}\,N_c\,(N_c^2-1)^2}\sum_{j,\bar{j}=\gamma^\ast,Z}\frac{\tilde{C}_j\tilde{C}_{\bar{j}}\,(a_j a_{\bar{j}}+b_j b_{\bar{j}})}{(M_B^2-\Delta_j)\,(M_B^2-\Delta_{\bar{j}})^\ast}~.
\end{equation}
The advantage of the quantities $N^{(i)}$ is that they are $\--$ in principle $\--$ Lorentz-invariant after having integrated out the CS-angles. Hence, they can be evaluated directly in the hadron c.m.-frame, i.e., the lab frame at the LHC.

Below we use these formulae to estimate the size of the effects from linearly polarised gluons that one can expect at the LHC by measuring the associated heavy-quarkonium + $Z$ - final state.

\section{Numerical Predictions for an associated quarkonium + $Z$ - final state}
\label{numericsZ}

We can numerically compare the relative size of both contributions from unpolarised and linearly polarised gluons to the angular-integrated cross section $N^{(0)}$ in (\ref{WeightedStructures}) by considering the LO ratios  $\hat{F}_{2,3,4}(Q,\alpha,\beta)/\hat{F}_1(Q,\alpha,\beta)$ from Eqs.~(\ref{F1int} -  \ref{F4int}). In this section, we focus on the production of a (quasi)-real $Z$-boson, i.e., dileptons with an invariant mass around the $Z$-pole mass $M_B\simeq m_Z$. To this end, we consider a $M_B$-bin around $m_Z$ with a bin size of, say, 2 GeV. Hence, we {\it define} cross sections for real $Z$-boson production in the following way,
\begin{equation}
\frac{\mathrm{d}\sigma^{pp\to J/\psi[\Upsilon]ZX}}{\mathrm{d}^4q}\equiv \int_{m_Z-1\,\mathrm{GeV}}^{m_Z+1\,\mathrm{GeV}}\mathrm{d}M_B\,\frac{\mathrm{d}\sigma^{pp\to J/\psi[\Upsilon]\ell \bar{\ell}X}}{\mathrm{d}^4q\,\mathrm{d}M_B^2}.\label{ZProd}
\end{equation}
The structures of the corresponding quantities $N^{(i)}$ in (\ref{WeightedStructures}) remain valid for real $Z$-boson production, but with integrated partonic prefactors
\begin{equation}
\hat{F}_i^Z(Q,\alpha) = 2Q^2 \int_{\frac{m_Z-1\,\mathrm{GeV}}{Q}}^{\frac{m_Z+1\,\mathrm{GeV}}{Q}}\mathrm{d}\beta\,\beta\,\hat{F}_i(Q,\alpha,\beta)\,.\label{FZ}
\end{equation}

\begin{figure}[t]
\centering
\subfloat[]{\includegraphics[width=0.49\textwidth,angle=0]{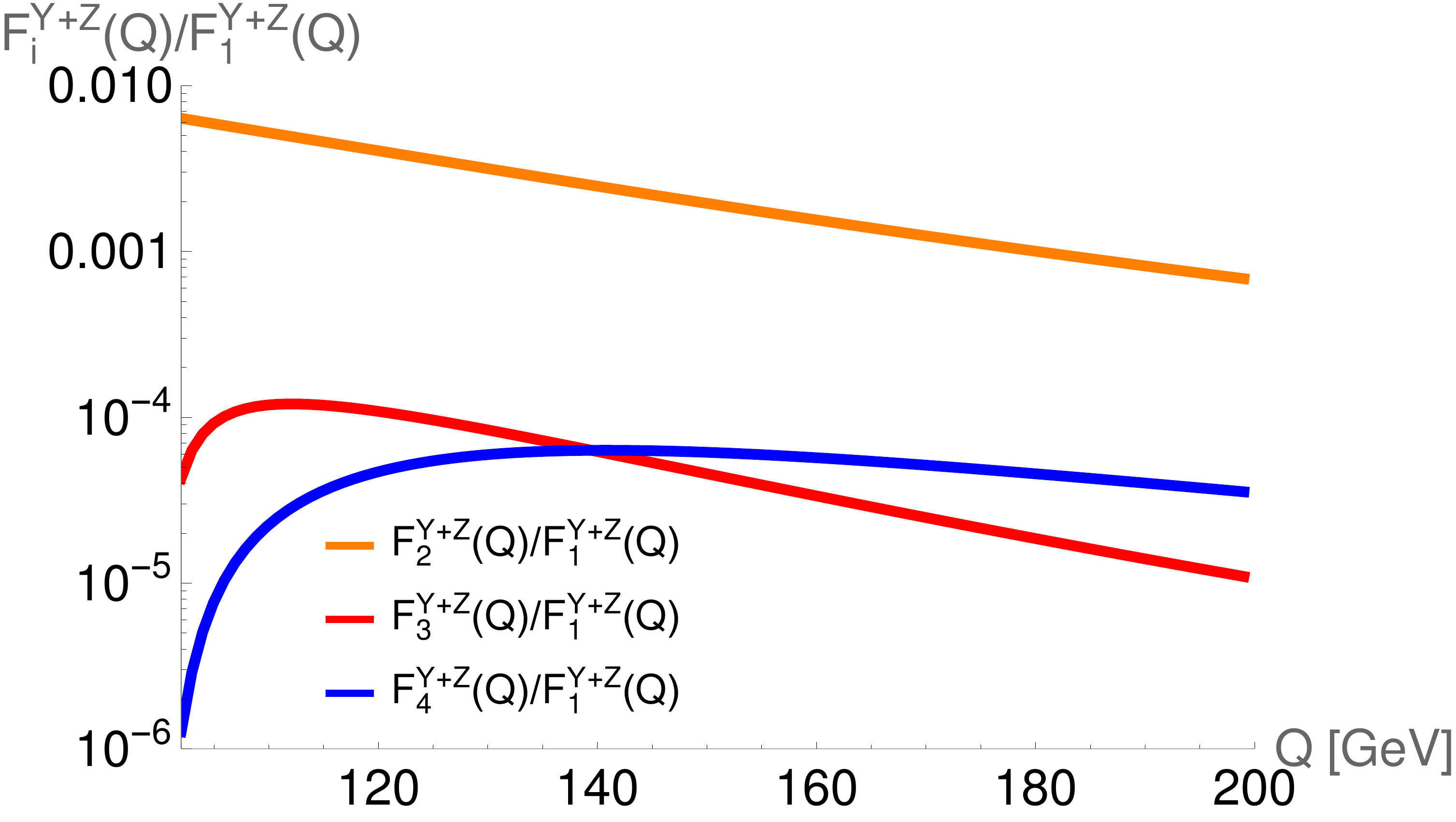}}
\subfloat[]{\includegraphics[width=0.49\textwidth,angle=0]{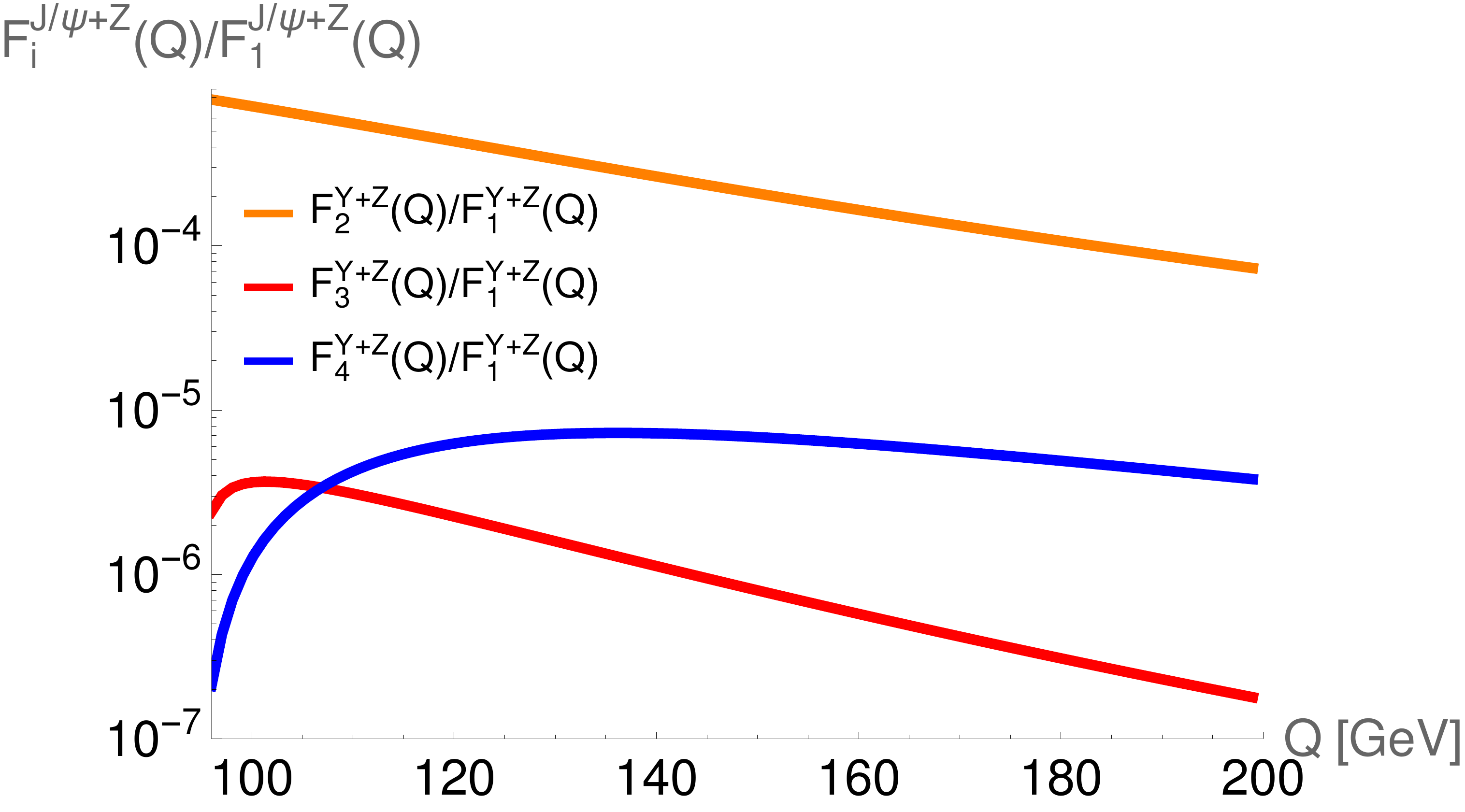}}
\caption{The ratios $\hat{F}_{2,3,4}^Z(Q)/\hat{F}_1^Z(Q)$ from (\ref{FZ}) plotted vs. the invariant final-state mass $Q\ge M_{\cal Q}+m_Z$ for a $\Upsilon$ (a) and a $J/\psi$ (b).\label{fig:FiF1Z}}
\end{figure}

The LO ratios $\hat{F}_{2,3,4}(Q,\alpha,\beta)/\hat{F}_1(Q,\alpha,\beta)$ for real $Z$-boson productions are shown in Fig.~\ref{fig:FiF1Z}. In the left plot we present our result for an associated $\Upsilon$ state with mass  $m_\Upsilon=9.46\,\mathrm{GeV}$. In the right plot, we have shown it for a $J/\psi$ state with mass $m_{J/\psi}=3.1\,\mathrm{GeV}$. First of all, we observe that the ratios are rather small in general: the ratio $F_2/F_1$ is about half of a percent at most for $\Upsilon$-production and even smaller ($\le 10^{-3}$) for $J/\psi$ production.  One can expect that the convolution  $\mathcal{C}[w_2\, h_1^{\perp g}\, h_1^{\perp g}]$ in (\ref{WeightedStructures}) from linearly polarised gluons does not exceed in size the convolution $\mathcal{C}[f_1^g\, f_1^g]$ \cite{Boer:2011kf,Boer:2014tka}. For example, the models of Ref.~\cite{Boer:2014tka} indicate that the (scale-dependent) ratio $R=\mathcal{C}[w_2\, h_1^{\perp g}\, h_1^{\perp g}]/\mathcal{C}[f_1^g\, f_1^g]$ is {\it at most} about $2/3$ for a small scale $Q\sim 3\,\mathrm{GeV}$, but typically (much) smaller for larger scales. Hence, it is not unreasonable to neglect the contribution from linearly polarised gluons for the quantity $N^{(0)}$ in (\ref{WeightedStructures}) and approximate to good accuracy,
\begin{equation}
N^{(0)}=\frac{\mathrm{d}\sigma^{pp\to J/\psi[\Upsilon]\ell \bar{\ell}X}_{\mathrm{TMD,\,LO}}}{\mathrm{d}^4q\,\mathrm{d}M_B^2}\simeq \hat{F}_1(Q,\alpha,\beta)\,\mathcal{C}[f_1^g\,f_1^g].\label{ApprN0}
\end{equation}

In a next step we evaluate the quantities $N_Z^{(i)}$ for real $Z$-boson production in the hadron c.m.-frame where $P^\mu_{a/b}=(\sqrt{S}/2)(1,0,0,\pm 1)$ and $q^\mu=(\sqrt{Q^2+{\bm q}_T^2}\,\cosh Y,{\bm q}_T,\sqrt{Q^2+{\bm q}_T^2}\,\sinh Y)$. Here, $Y$ denotes the rapidity of quarkonium-dilepton pair, i.e., the rapidity of the final state. Also, $\mathrm{d}^4q=Q\,\mathrm{d}Q\,\mathrm{d}Y\,\mathrm{d}^2{\bm q}_T$. We then investigate the following distributions (ratios) that where already proposed in Ref.~\cite{Dunnen:2014eta},
\begin{eqnarray}
S_Z^{(0)}(Q,Y,{\bm q}_T)\equiv \frac{N_Z^{(0)}}{\int_0^{Q^2/4}\mathrm{d}{\bm q}_T^2\,N_Z^{(0)}} &=&\frac{\mathcal{C}[f_1^g\, f_1^g]}{\int_0^{Q^2/4}\mathrm{d}{\bm q}_T^2\,\mathcal{C}[f_1^g\,f_1^g]}\,,\label{S0}\\
S_Z^{(2)}(Q,Y,{\bm q}_T)\equiv \frac{N_Z^{(2)}}{\int_0^{Q^2/4}\mathrm{d}{\bm q}_T^2\,N_Z^{(0)}} &=&\frac{\hat{F}_3^Z(Q,\alpha)}{\hat{F}_1^Z(Q,\alpha)}\frac{\mathcal{C}[w_{3a}\,h_1^{\perp g}\,f_1^g]+\mathcal{C}[w_{3b}\,f_1^g\,h_1^{\perp g}]}{\int_0^{Q^2/4}\mathrm{d}{\bm q}_T^2\,\mathcal{C}[f_1^g\,f_1^g]}\,,\label{S2}\\
S_Z^{(4)}(Q,Y,{\bm q}_T)\equiv \frac{N_Z^{(4)}}{\int_0^{Q^2/4}\mathrm{d}{\bm q}_T^2\,N_Z^{(0)}} &=&\frac{\hat{F}_4^Z(Q,\alpha)}{\hat{F}_1^Z(Q,\alpha)}\frac{\mathcal{C}[w_{4}\,h_1^{\perp g}\,h_1^{\perp g}]}{\int_0^{Q^2/4}\mathrm{d}{\bm q}_T^2\,\mathcal{C}[f_1^g\,f_1^g]}\,.\label{S4}
\end{eqnarray}
Note that, for these ratios, the transverse momentum of the final state ${\bm q}_T$ $\--$ the quarkonium-dilepton pair's transverse momentum imbalance $\--$ is restricted to be smaller than $Q/2$ in order to roughtly fulfill the TMD factorisation requirement $|{\bm q}_T|\ll Q$.

\begin{figure}[t]
\centering
\subfloat[]{\includegraphics[width=0.32\textwidth,angle=0]{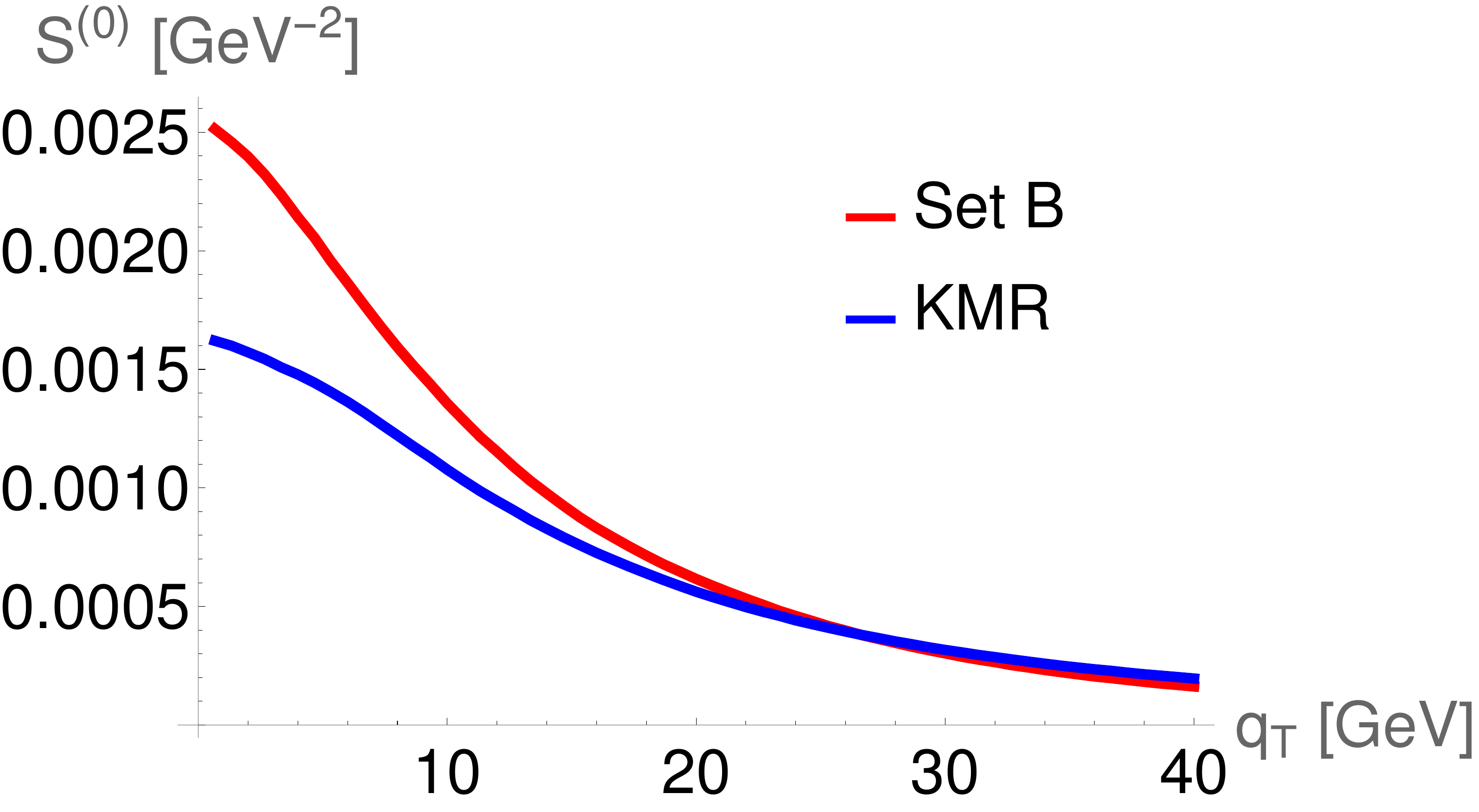}\hspace*{0.2cm}}
\subfloat[]{\includegraphics[width=0.32\textwidth,angle=0]{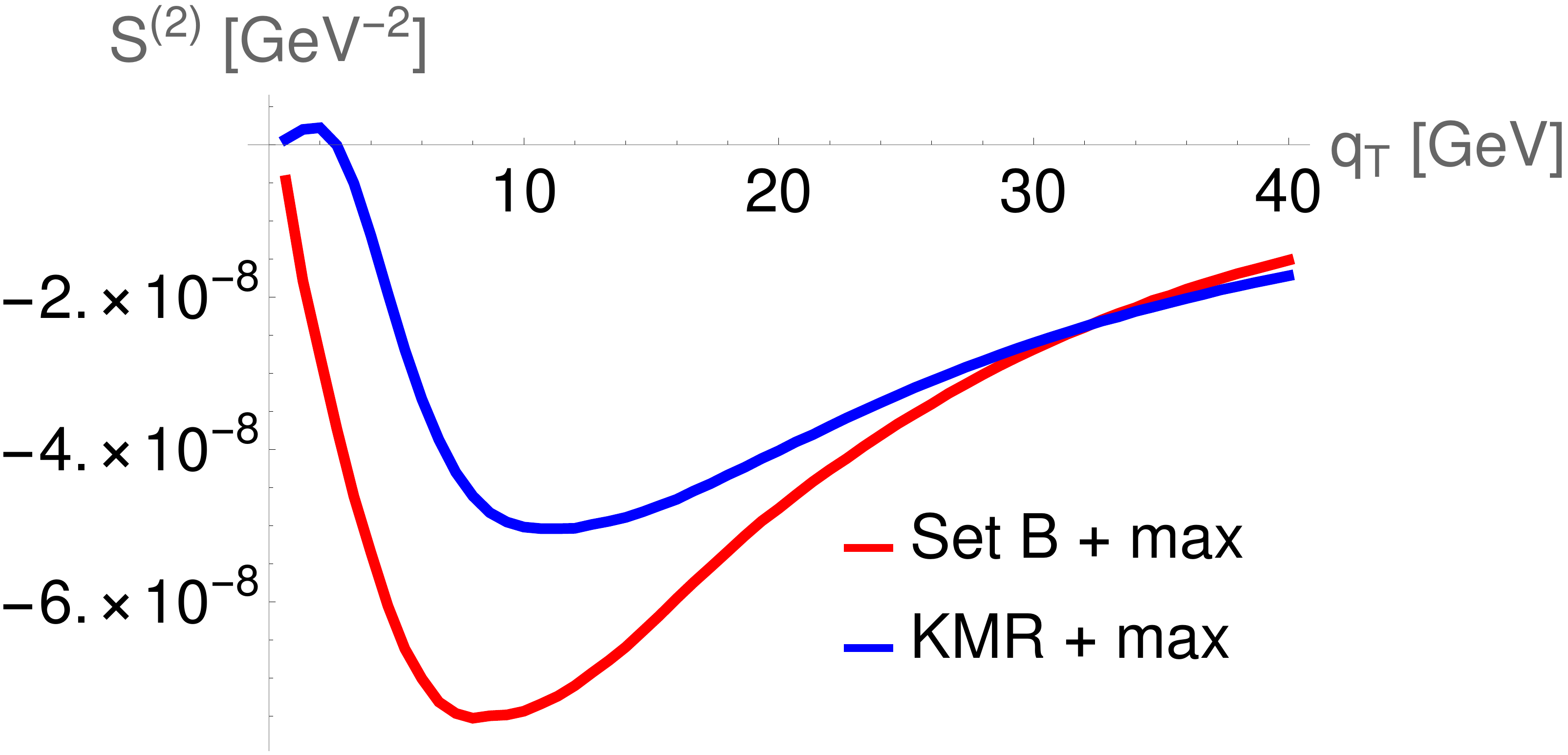}\hspace*{0.2cm}}
\subfloat[]{\includegraphics[width=0.32\textwidth,angle=0]{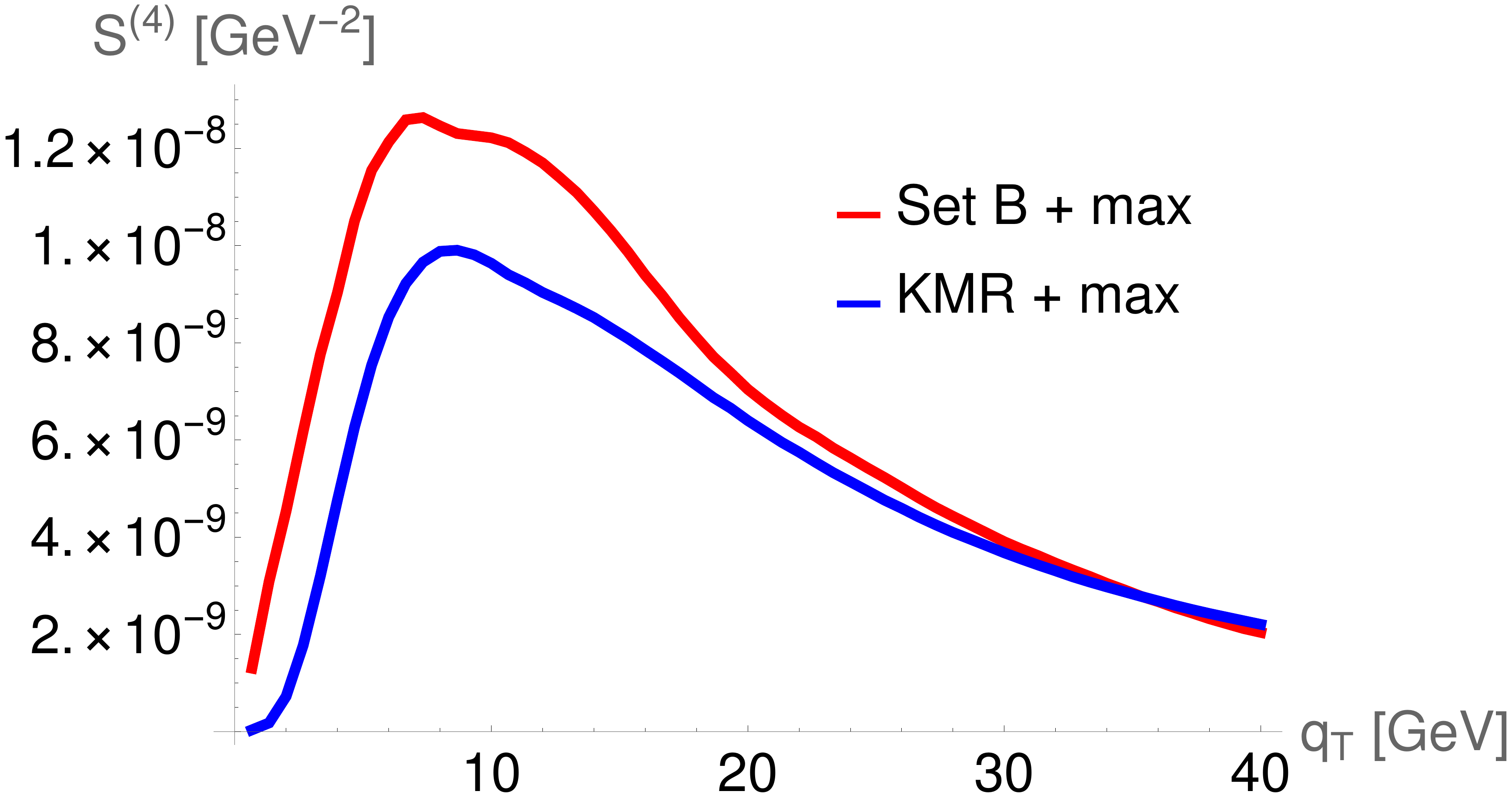}}
\caption{The azimuthally independent distribution $S^{(0)}$ (a) and the azimuthal distributions $S^{(2)}$ (b) and $S^{(4)}$ (c) at the LHC for $\sqrt{S}=14\,\mathrm{TeV}$ for the $\Upsilon + Z$ final state at midrapidity and an invariant mass $Q=120\,\mathrm{GeV}$.\label{fig:SZ}}
\label{fig:Sz}
\end{figure}

In order to estimate the size of the effects of the linearly polarised gluons hidden in the observables $S_Z^{(2)}$ and $S_Z^{(4)}$, it is instructive to first calculate the ratios of the perturbative prefactors $\hat{F}_3^Z/\hat{F}_1^Z$ and $\hat{F}_4^Z/\hat{F}_1^Z$. We show these ratios in Fig.~\ref{fig:FiF1Z} as well, plotted vs. the invariant mass $Q$ for a $\Upsilon$ and $J/\psi$ state. We find rather small ratios, around $10^{-4}$ for $\hat{F}_3^Z/\hat{F}_1^Z$  and $5\times 10^{-5}$ for $\hat{F}_4^Z/\hat{F}_1^Z$ at the peak around $Q=120\,\mathrm{GeV}$ for a $\Upsilon$-particle. The ratios for $J/\psi$-production are even smaller, of the order of $10^{-6}$. This is in contrast to the $\Upsilon+\gamma$ final state discussed in Ref.~\cite{Dunnen:2014eta} where the corresponding ratios are about $0.05$ and $0.03$, respectively, for an invariant mass $Q=20\,\mathrm{GeV}$. This already suggests that the final state containing a heavy quarkonium state and a real $Z$-boson may not be sufficiently suitable to study the distribution of linearly polarised gluons, $h_1^{\perp g}$.
A more quantitative statement about the feasibility of measurements of the azimuthal observables $S^{(2)}$ and $S^{(4)}$ can be given through an estimate of the convolution integrals in Eqs.~(\ref{S0},\ref{S2},\ref{S4}). We use the same model {\it Ans\"atze} for the unpolarised gluon $f_1^g$ that were also adopted in the predictions for a $\Upsilon + \gamma$ state in Ref.~\cite{Dunnen:2014eta}, where two parameterisations of the {\it unintegrated gluon distribution} (UGD) have been used as an input for the TMD gluon function $f_1^g$. Although the UGD established for small-$x$ physics may not be identified in general with the unpolarised gluon TMD, such an input serves as a first numerical estimate of the size that can be expected for the distributions $S^{(i)}$. In particular, we use the Set B0 solution to the CCFM equation with an initial condition based on the HERA data from Refs.~\cite{Jung:2004gs,Jung:2010si} and the KMR parameterisation from Ref.~\cite{Kimber:2001sc} for the UGD. 
In Ref.~\cite{Dunnen:2014eta} the saturation of the positivity bound \cite{Mulders:2000sh} for the distribution of linearly polarised gluons was assumed, i.e., $h_1^{\perp g}(x,{\bm k}_T^2)=2M^2/{\bm k}_T^2\,f_1^g(x,{\bm k}_T^2)$. We rely on the same assumption in this work as well. 
In Fig.~\ref{fig:SZ} the azimuthal ${\bm q}_T$-distribution $S^{(2)}$ and $S^{(4)}$ from Eqs.~(\ref{S2},\ref{S4}) are shown for real $Z$-bosons and are of the size of about $10^{-8}$, which is about four orders of magnitudes smaller than the corresponding contributions for a $\Upsilon + \gamma$ final state. The ${\bm q}_T$-integrated azimuthal observables $\int d^2{\bm q}_T\,S^{(2)}$ and $\int d^2{\bm q}_T\,S^{(4)}$ amount to roughly $0.007\%$ and $0.001\%$, respectively $\--$ three orders of magnitude smaller than for $\Upsilon + \gamma$. Hence, we conclude that it will be very difficult to access the linearly polarised gluons with a $\Upsilon + Z$ final state. Having said that, this particular final state might be a suitable candidate for studying experimentally the unpolarised gluon TMD on its own through the measurement of the azimuthally independent distribution $S^{(0)}$. The difference between this observable for a $\Upsilon + Z$ final state and a $\Upsilon + \gamma$ final state is that the invariant mass varies. In Ref.~\cite{Dunnen:2014eta} the distribution $S^{(0)}$ was studied at $Q=20\,\mathrm{GeV}$ for a $\Upsilon + \gamma$ final state, while in Fig.~\ref{fig:SZ} this distribution is shown for $Q=120\,\mathrm{GeV}$. The larger invariant mass is a consequence of the mass of the $Z$-boson. Hence, the ${\bm q}_T$-distribution $S^{(0)}$ is much broader for $\Upsilon + Z$. In addition, the detection of a dilepton pair at the $Z$-pole is experimentally much favorable in contrast to real photon detection because an isolation procedure is needed in the latter case.

\section{Numerical prediction for quarkonium + a dilepton from an off-shell photon}
\label{numericsDY}

\begin{figure}[t]
\centering
\subfloat[]{\includegraphics[width=0.49\textwidth,angle=0]{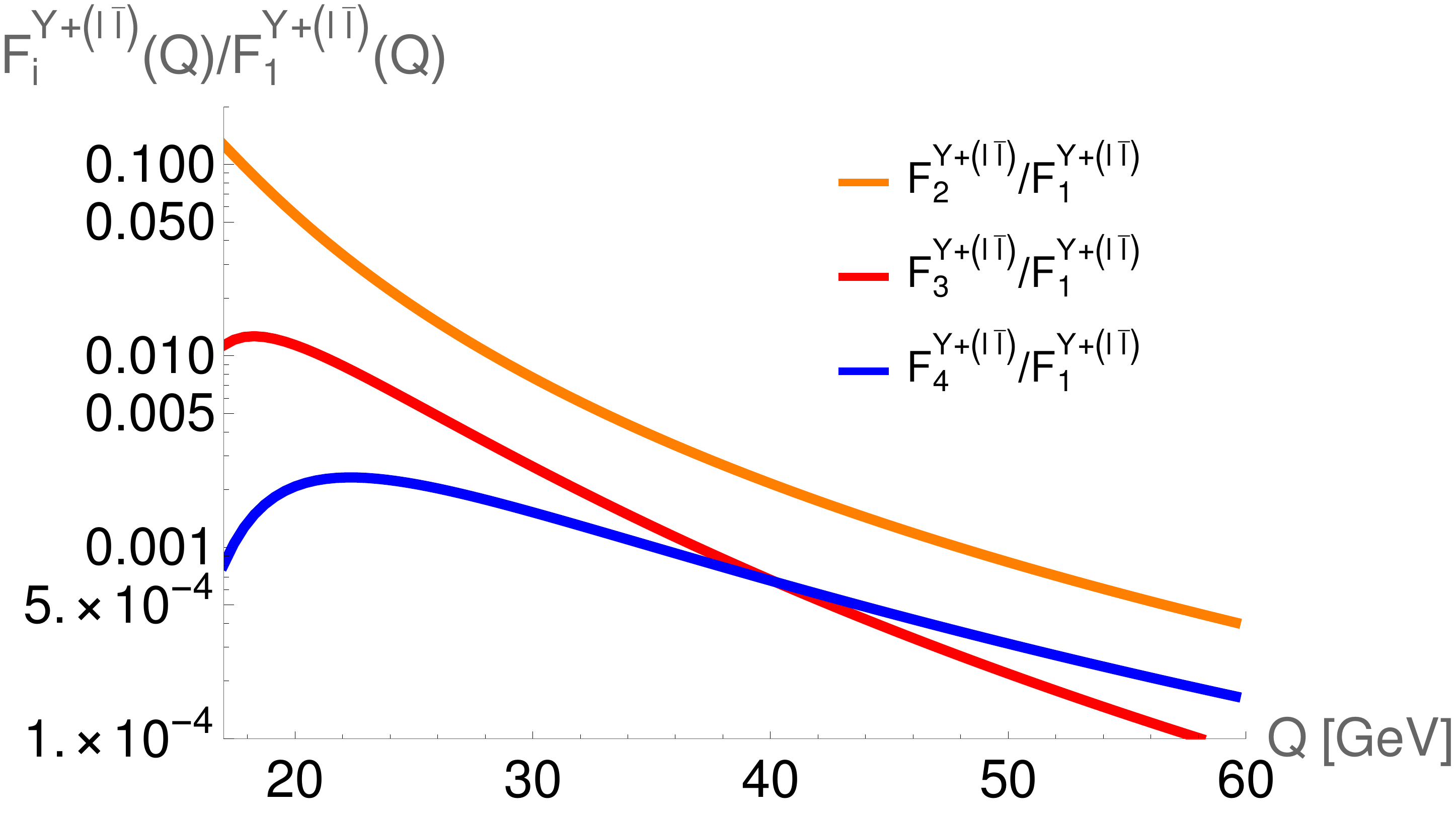}}
\subfloat[]{\includegraphics[width=0.49\textwidth,angle=0]{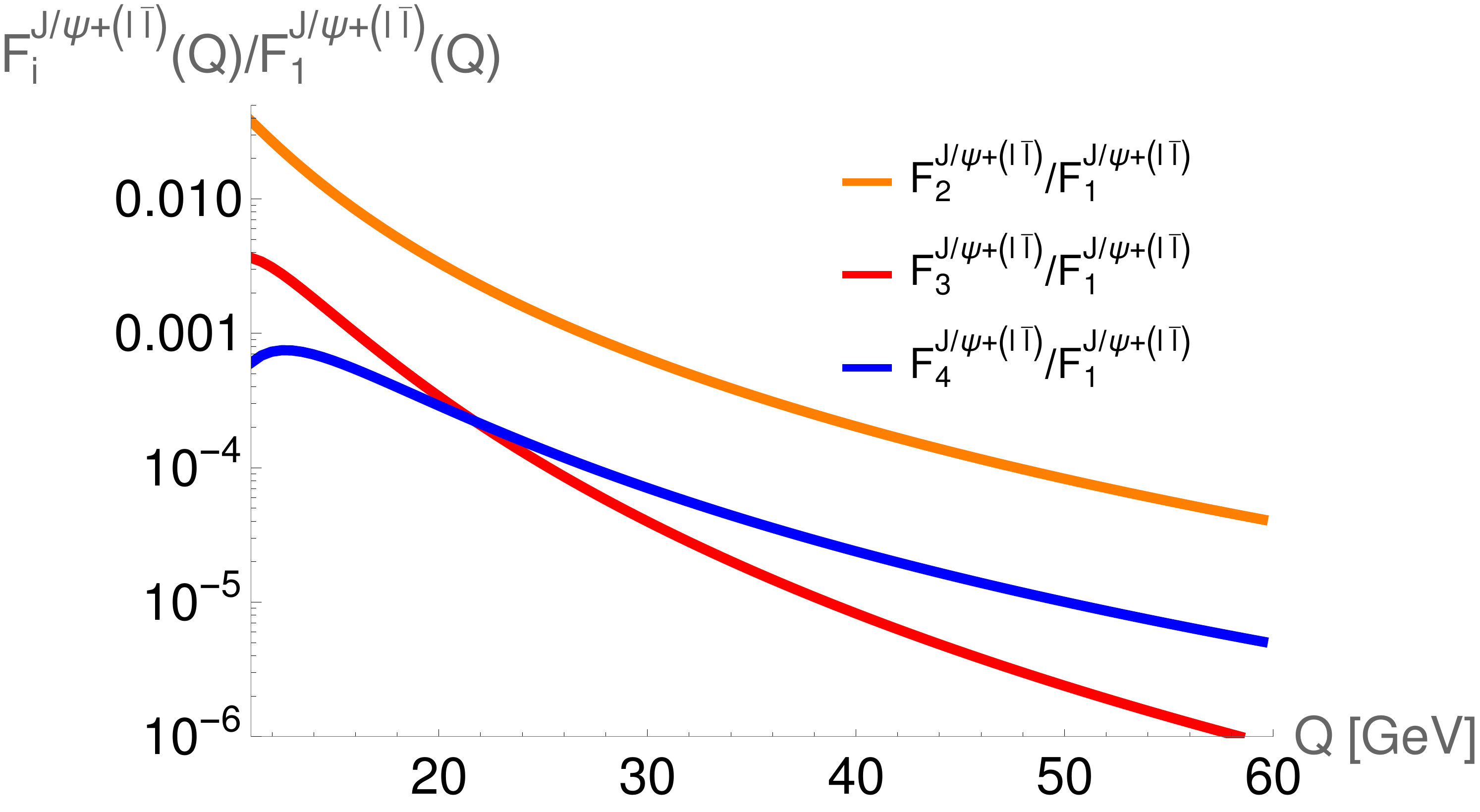}}
\caption{The ratios $\hat{F}_i^{\ell \bar{\ell}}(Q)/\hat{F}_1^{\ell \bar{\ell}}(Q)$ and $\hat{F}_4^Z(Q)/\hat{F}_1^Z(Q)$ from (\ref{FZ}) plotted vs.\ the invariant final-state mass $Q$ for a $\Upsilon$ (a) and a $J/\psi$ (b). The lepton pair is selected such that only small invariant masses $M_B\in[5\,\mathrm{GeV},7\,\mathrm{GeV}]$ are allowed.\label{fig:FiF1MB5}}
\end{figure}

In this section we repeat the steps of the previous section but we consider lepton pairs with a relatively small invariant mass $M_B\in [5\,\mathrm{GeV},\,7\,\mathrm{GeV}]$ between the $J/\psi$ and $\Upsilon$ families. This mass range is far away from the $Z$-pole mass $m_Z$, and dilepton creation from decays of virtual photons instead of $Z$-bosons should dominate. 

\begin{figure}[b]
\centering
\subfloat[]{\includegraphics[width=0.32\textwidth,angle=0]{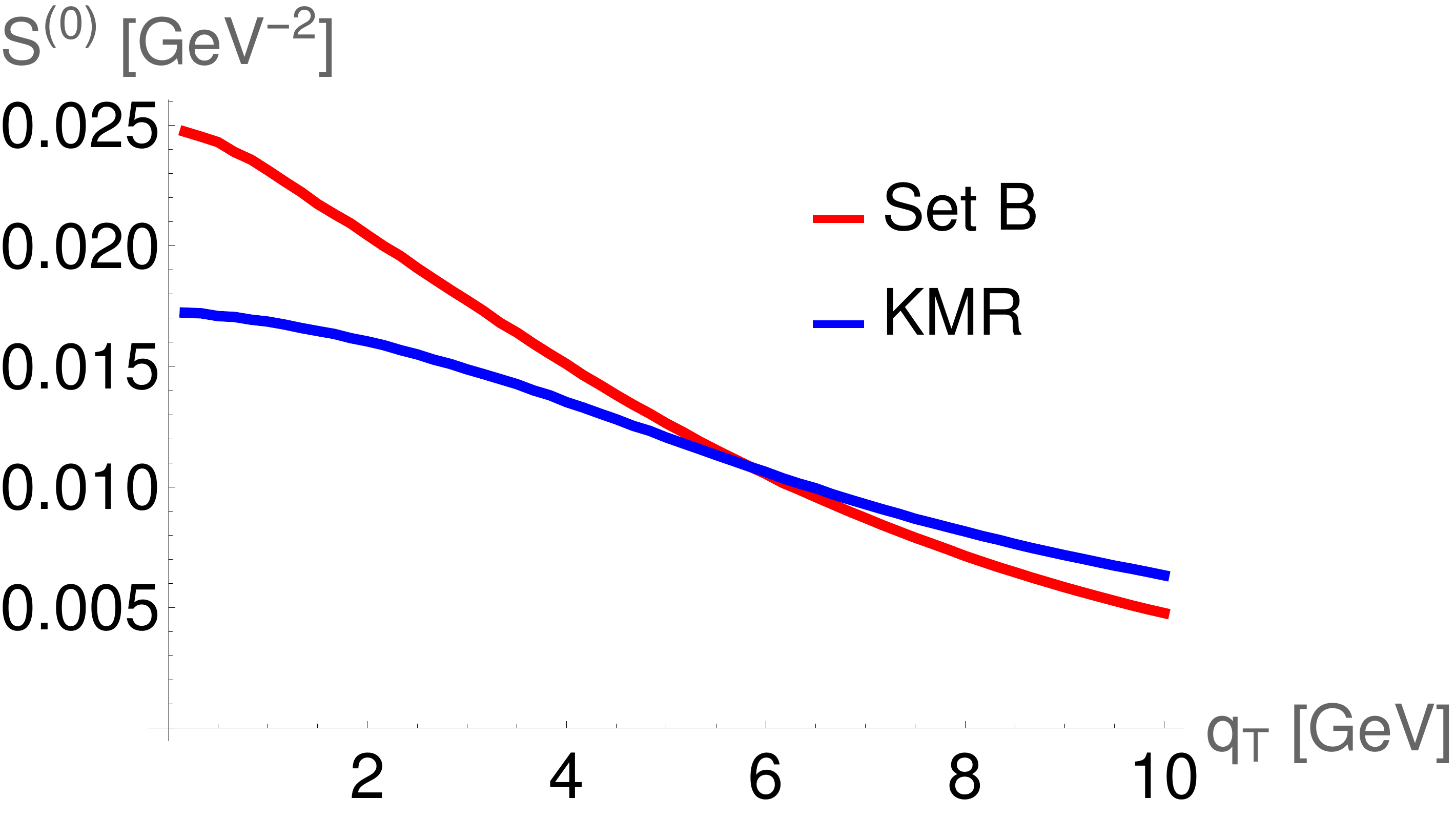}\hspace*{0.2cm}}
\subfloat[]{\includegraphics[width=0.32\textwidth,angle=0]{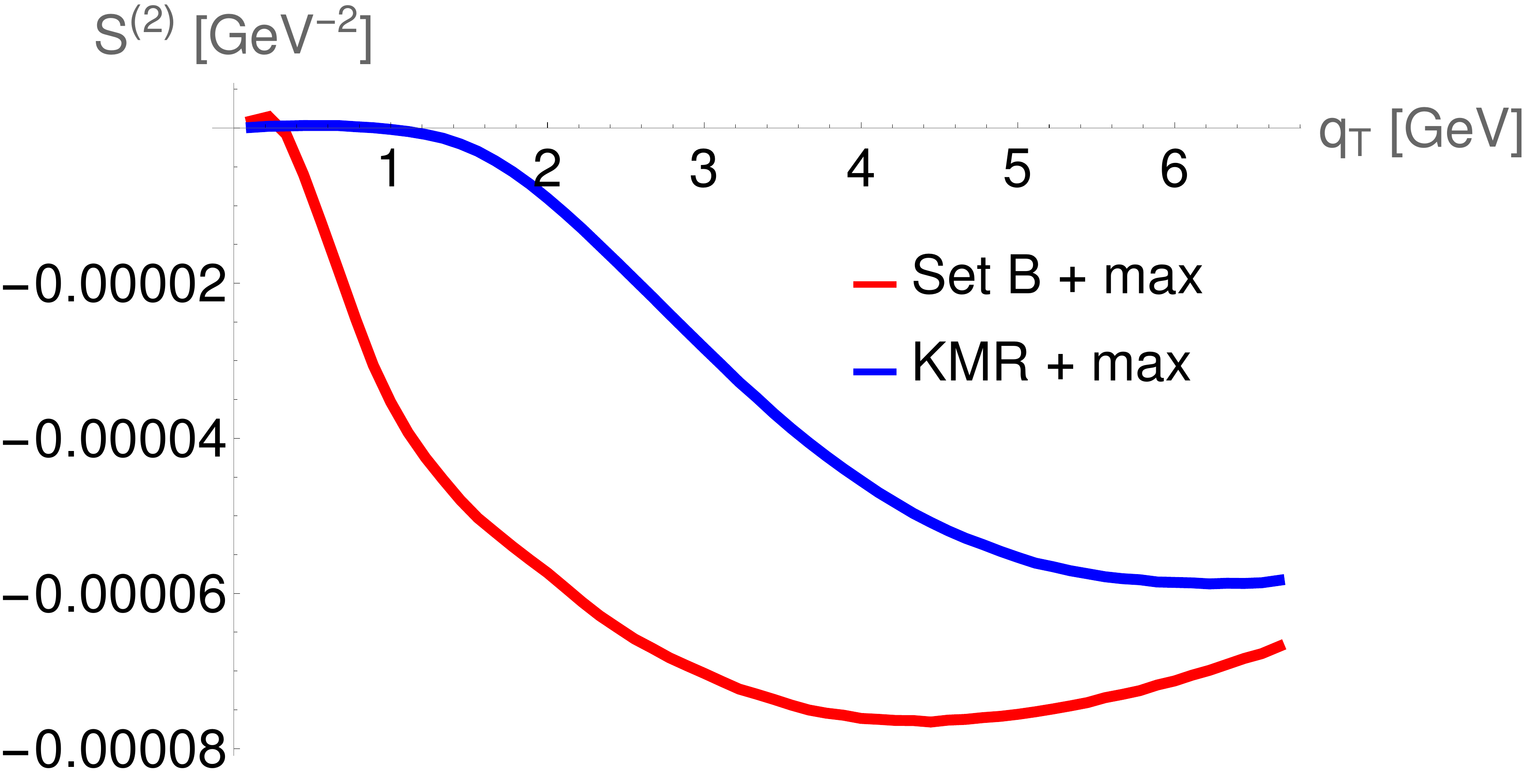}\hspace*{0.2cm}}
\subfloat[]{\includegraphics[width=0.32\textwidth,angle=0]{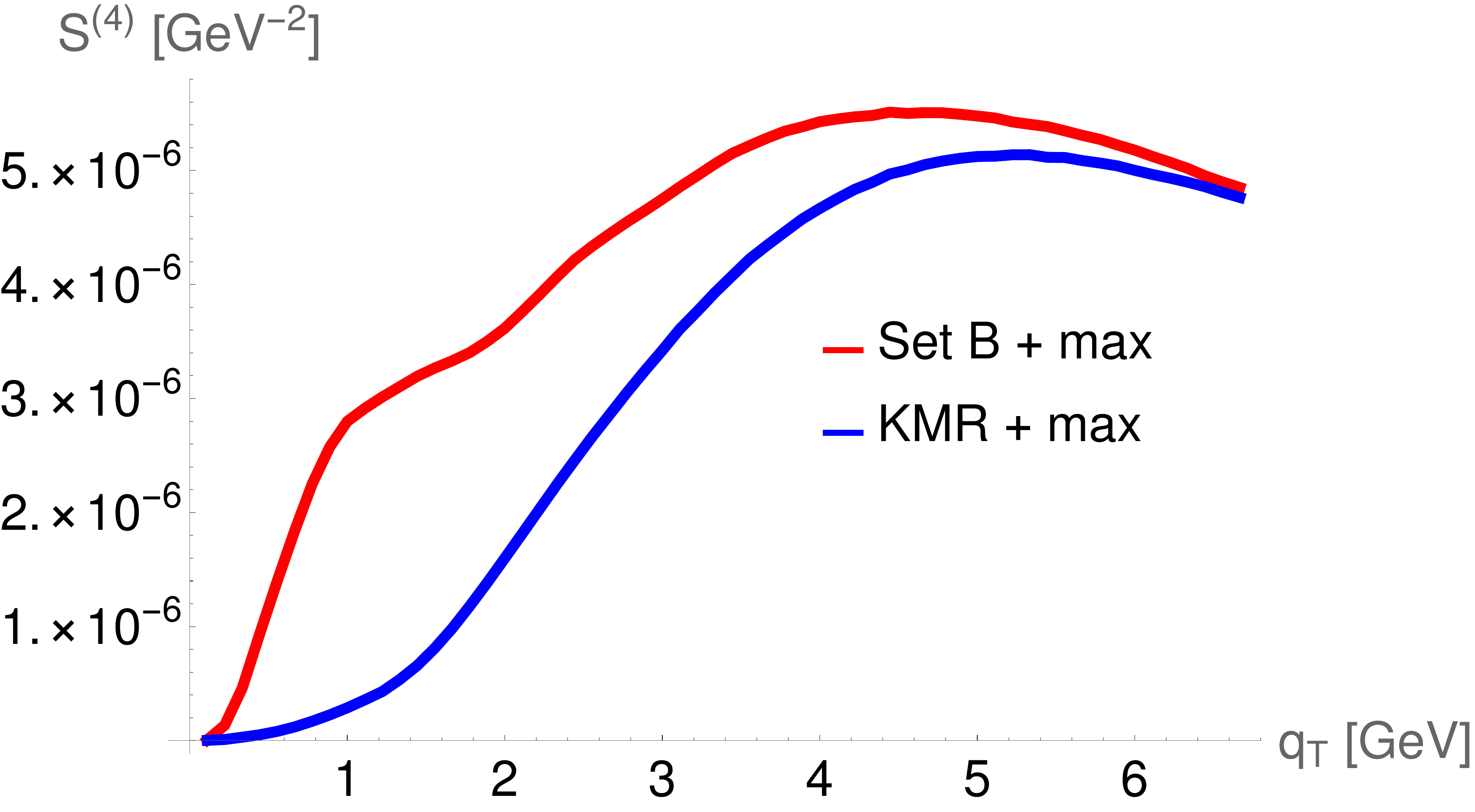}}
\caption{The azimuthally independent distribution $S^{(0)}$ (a) and the azimuthal distributions $S^{(2)}$ (b) and $S^{(4)}$ (c) for a $\Upsilon + (\ell\bar{\ell})$ final state at midrapidity and an invariant mass $Q=20\,\mathrm{GeV}$. The dilepton mass range is $M_B\in [5\,\mathrm{GeV},\,7\,\mathrm{GeV}]$.\label{fig:SMB5}}
\end{figure}

The advantage is that we can investigate lower values of $Q$ with a minimum final state invariant mass $Q_{\mathrm{min}}=7\,\mathrm{GeV}+M_{\cal Q}$. In fact, we expect to observe some similarities with the associated $\mathcal{Q}+\gamma$ final state discussed in Ref.~\cite{Dunnen:2014eta}. The theoretical formulae for this kinematic range are the same as for real $Z$-boson production of the last section, except that the integration region in (\ref{FZ}) is $5\,\mathrm{GeV}\le M_B\le 7\,\mathrm{GeV}$.

In Fig. \ref{fig:FiF1MB5}, we plot the ratios $F_i/F_1$ for associated Quarkonium - dilepton production with small dilepton masses vs. the final state invariant mass $Q$. We observe that these ratios are considerably larger (by  1-2 orders of magnitude) than for real $Z$-boson production, but still smaller than for real photon production. Overall, one can say that the associated Quarkonium - dilepton final state is rather sensitive to the dilepton mass $M_B$.  In particular, the ratio $F_2/F_1$ indicates that the prefactor $\hat{F_2}$ characterising the linearly polarised gluons may be a few percent of the factor $\hat{F}_1$ for lower $Q$. Nevertheless, we still consider the approximation (\ref{ApprN0}) to be justified.

We then calculate the $\bm{q}_T$-distributions $S^{(i)}$ for small dilepton masses and show our results in Fig.~\ref{fig:SMB5}. Since the distribution $S^{(0)}$ in the upper panel does not depend on the specific final state, we obtain the same result for a final state invariant mass $Q=20\,\mathrm{GeV}$ as for real photon production in Ref.~\cite{Dunnen:2014eta}. The ratios $F_{3,4}/F_1$ are smaller compared to real photon production as mentioned before, and this results in smaller azimuthal $\bm{q}_T$-distributions $S^{(2)}$ and $S^{(4)}$. We estimate the overall $\bm{q}_T$-integrated $\cos 2\phi $ effect from linearly polarised gluons to be about $0.5\%-0.6\%$ in the dilepton mass range $M_B\in [5\,\mathrm{GeV},\,7\,\mathrm{GeV}]$ for an invariant mass $Q=20\,\mathrm{GeV}$, while the $\cos 4\phi $ modulation amounts to $0.04\%-0.045\%$. Although these rates are smaller by an order of magnitude compared to real photon production, the experimental advantage of cleaner final state may outweigh the disadvantage of a smaller effect. This discussion is however beyond the scope of the present analysis.

\begin{figure}[t]
\centering
\subfloat[]{\includegraphics[width=0.49\textwidth,angle=0]{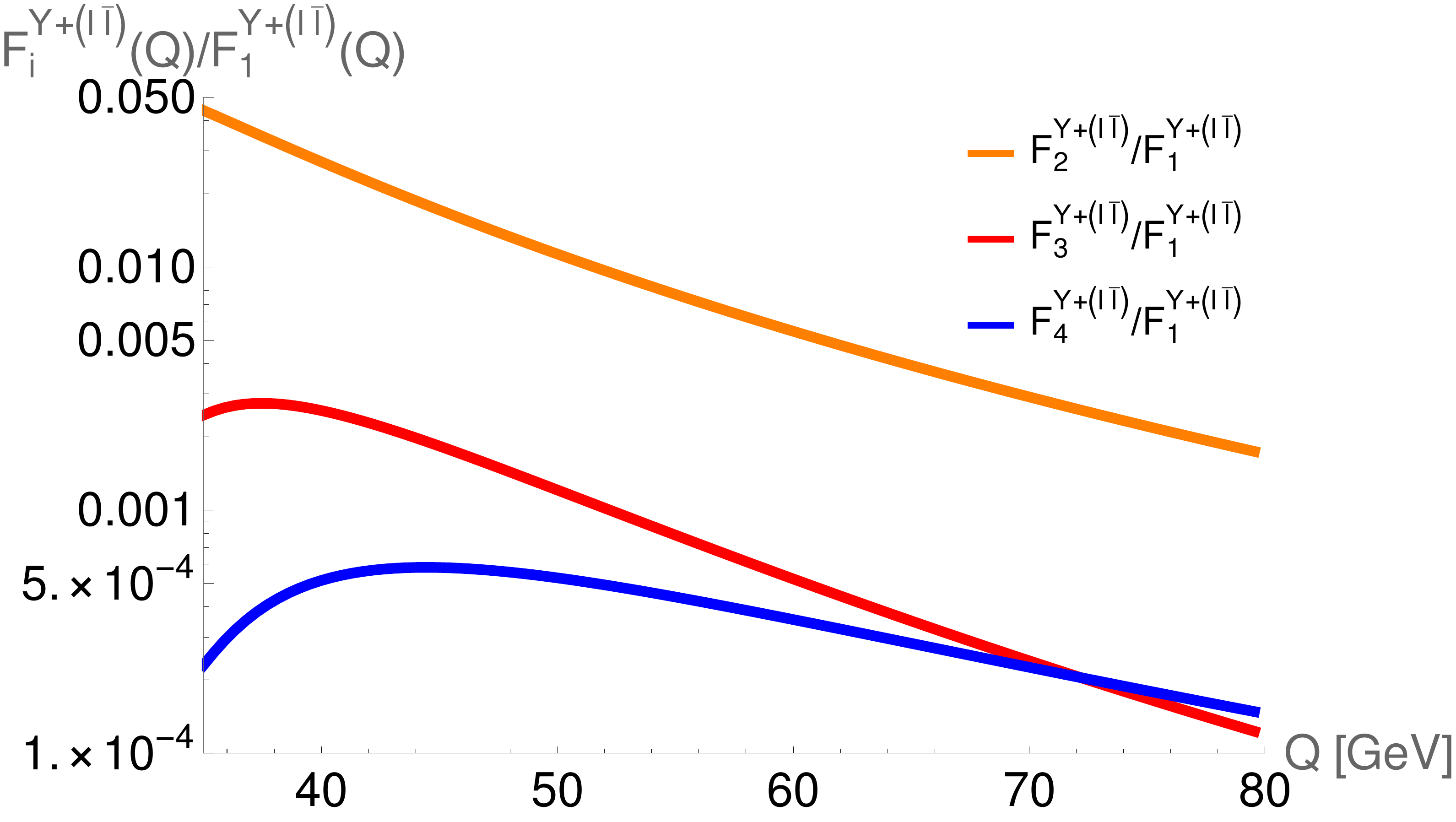}}
\subfloat[]{\includegraphics[width=0.49\textwidth,angle=0]{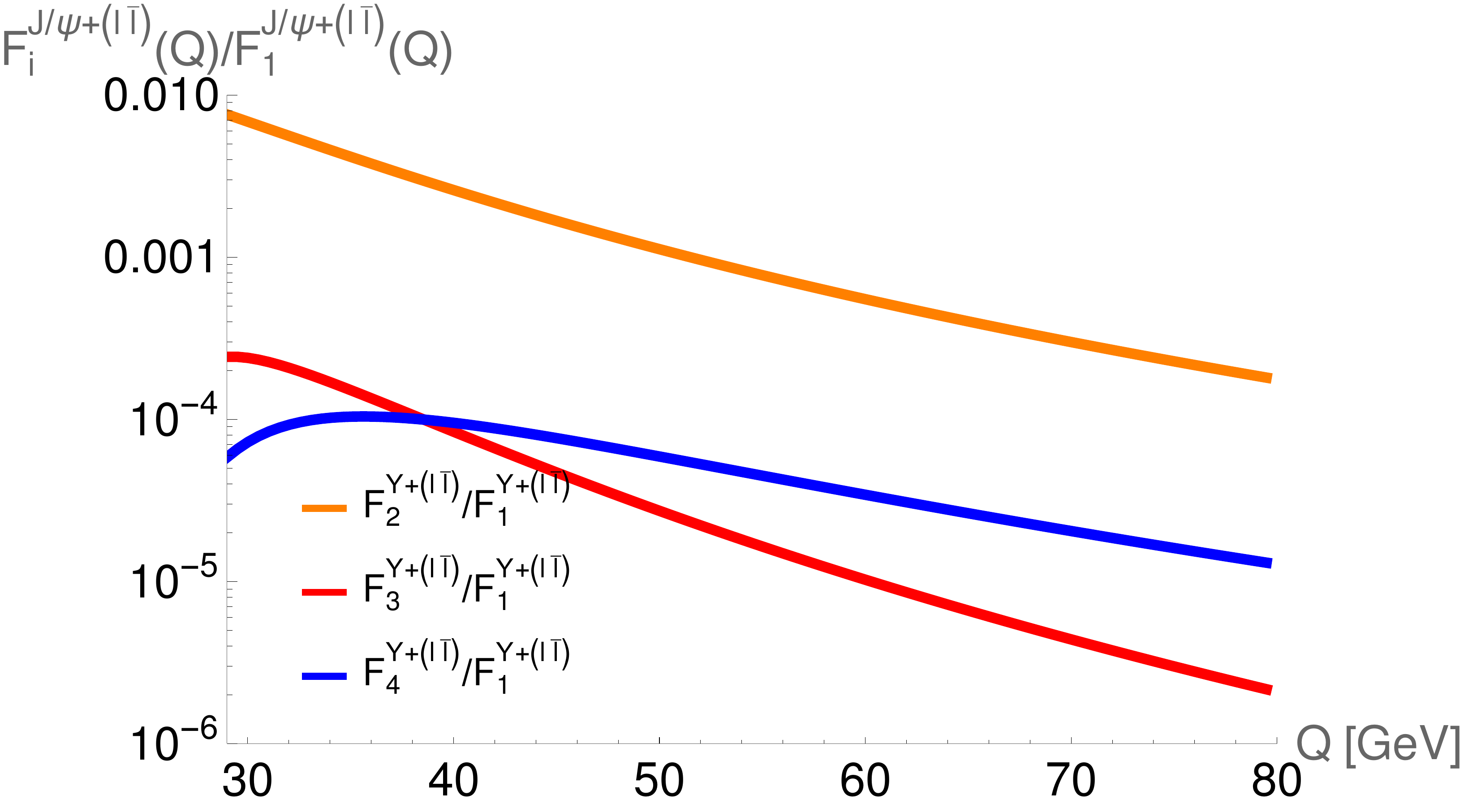}}
\caption{Same as Fig. \ref{fig:FiF1MB5}, but in the dilepton mass range $M_B\in[20\,\mathrm{GeV},25\,\mathrm{GeV}]$.\label{fig:FiF1MB20}}
\end{figure}

\begin{figure}[t]
\centering
\subfloat[]{\includegraphics[width=0.32\textwidth,angle=0]{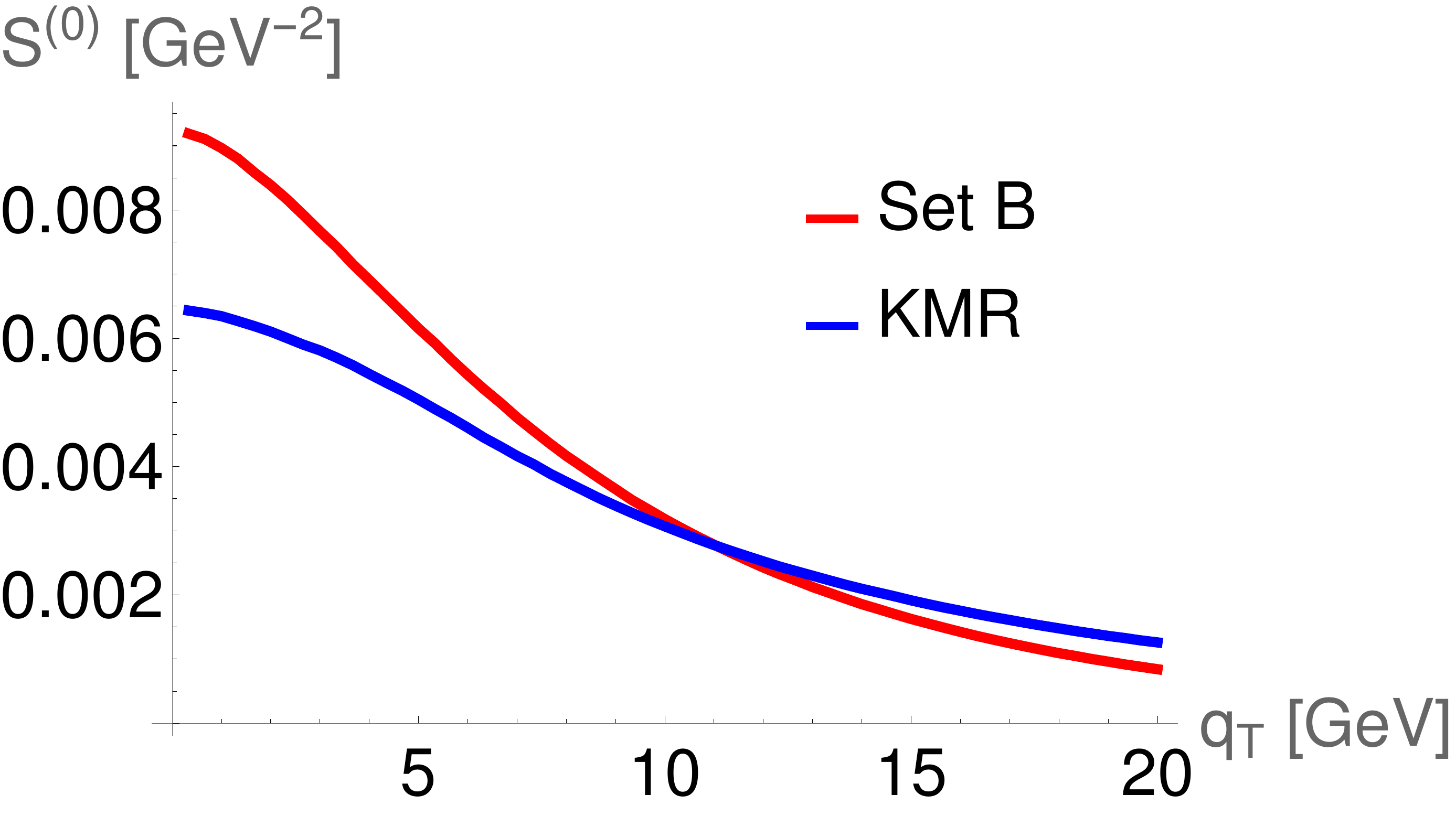}\hspace*{0.2cm}}
\subfloat[]{\includegraphics[width=0.32\textwidth,angle=0]{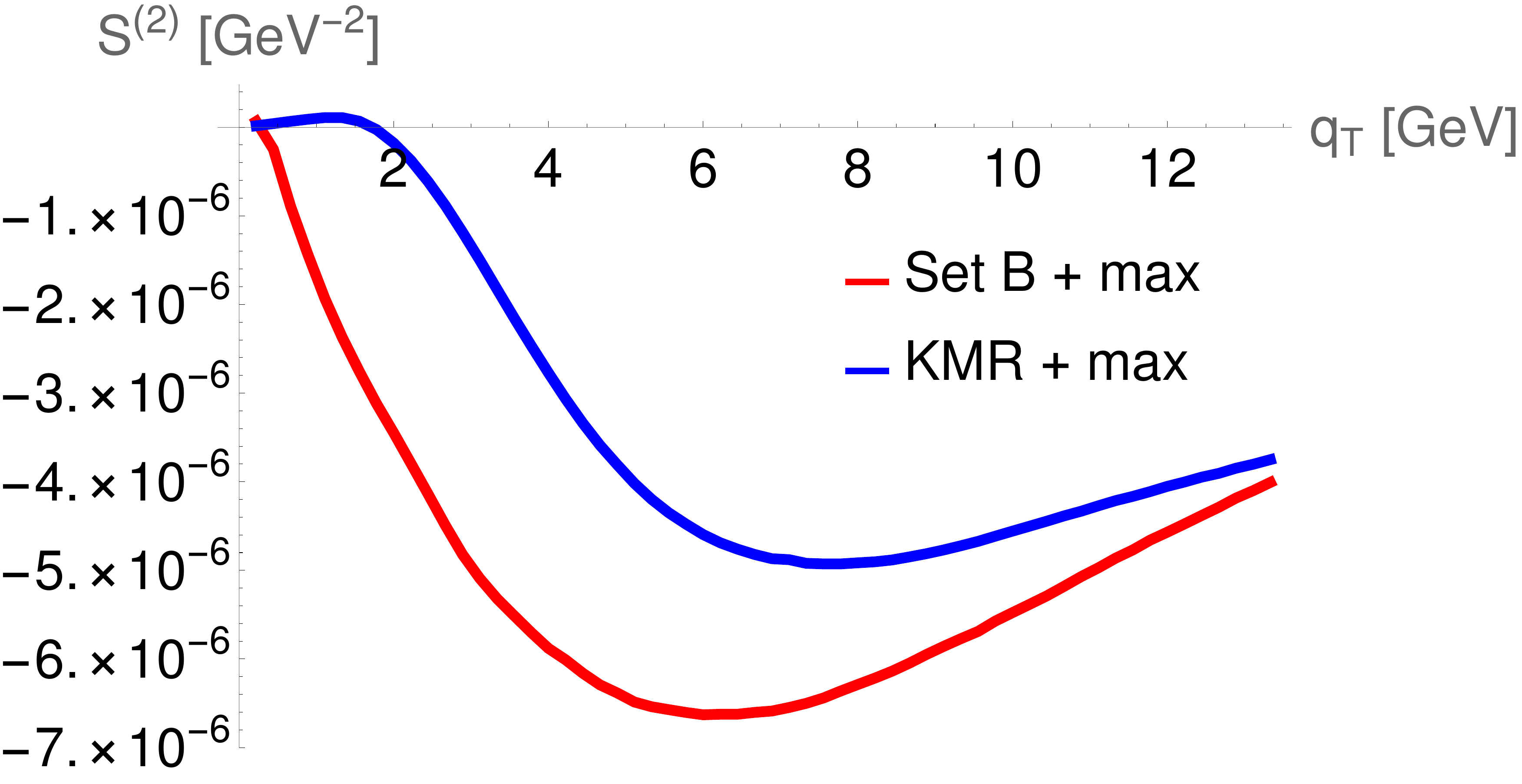}\hspace*{0.2cm}}
\subfloat[]{\includegraphics[width=0.32\textwidth,angle=0]{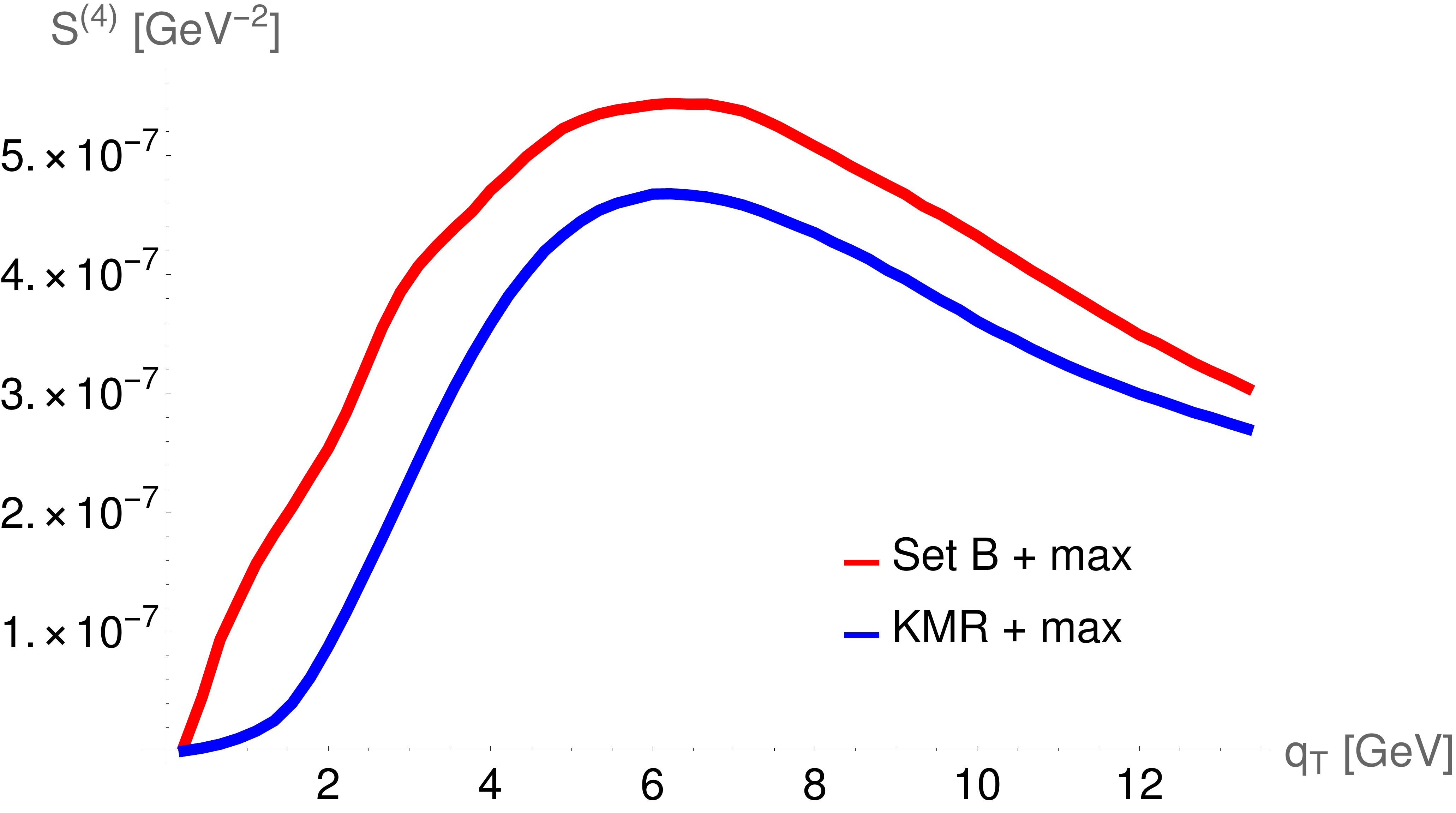}}
\caption{Same as Fig. \ref{fig:SMB5}, but in the dilepton mass range $M_B\in[20\,\mathrm{GeV},25\,\mathrm{GeV}]$.\label{fig:SMB20}}
\end{figure}

Finally, we also investigate an intermediate dilepton mass range $M_B\in[20\,\mathrm{GeV},25\,\mathrm{GeV}]$. The corresponding ratios of the LO prefactors $F_i/F_1$ are shown in Fig.~\ref{fig:FiF1MB20} where we observe a reduction of a factor of about 3-4 compared to the small dilepton mass range $M_B\in [5\,\mathrm{GeV},\,7\,\mathrm{GeV}]$, both for a $\Upsilon$ and $J/\psi$. Consequently, also the azimuthal $\bm{q}_T$-distributions $S^{(2)}$ and $S^{(4)}$, taken at a larger invariant final state mass $Q=40\,\mathrm{GeV}$ and shown in Fig.~\ref{fig:SMB20},  are smaller. The overall effect amounts to $0.13\%-0.15\%$ for the $\cos 2\phi$ modulation and $0.01\%-0.012\%$ for the $\cos 4\phi $ modulation.

\section{Conclusions}
\label{concl}

In this paper, we have presented the treatment of an arbitrary process induced by gluon-gluon fusion with a colour-singlet final state in the TMD approach. Using the helicity formalism, we have analysed the general structure of the fully differential, unpolarised cross section for this process in terms of TMD distributions of unpolarised and linearly polarised gluons inside an unpolarised nucleon. We have then calculated the partonic cross sections underlying the azimuthally independent and dependent structures of the hadronic cross section for a specific final state: a heavy quarkonium in a colour-singlet state and a real $Z$-boson. We have found that, in contrast to quarkonium production associated with a photon, the azimuthally dependent contributions are strongly suppressed with respect to the azimuthally independent ones. Therefore, the distribution of linearly polarised gluons in the nucleon is most probably not experimentally accessible for a quarkonium final state that is associated with a real $Z$-boson. The unpolarised gluon TMDs may however be studied by means of the azimuthally independent transverse momentum distribution of the process $pp\to \Upsilon Z X$, which can be directly measured at the LHC. 

We have also investigated the associated quarkonium + dilepton production for small and medium dilepton masses. These dileptons are predominantely generated from virtual photon decays. Although the effects from linearly polarised gluons are still smaller than for the real-photon case, a TMD extraction might be done for dilepton masses between the $J/\psi$ and $\Upsilon$ masses  at the LHC in view of the recent experimental studies of $\Upsilon+\Upsilon$~\cite{Khachatryan:2016ydm} and $J/\psi+\Upsilon$~\cite{Abazov:2015fbl} production.

\section*{Acknowledgements}
We thank D. Boer, W. den Dunnen, M.G. Echevarria, T. Kasemets, H.S. Shao, A. Signori, J.X. Wang for helpful discussions. The work of CP is supported by the European Research Council (ERC) under the European Union's Horizon 2020 research and innovation programme (grant agreement No.~647981, 3DSPIN). The work of JPL is supported in part
by the CNRS-IN2P3 (project TMD@NLO). The work of MS is supported in part by the Bundesministerium f\"ur Bildung und Forschung (BMBF) grant 05P15VTCA1.

\bibliographystyle{utphys}

\bibliography{psi-Z-TMD-300117}

\end{document}